\documentclass[aps,prd,amsmath,floats,amssymb,floatfix,twocolumn,superscriptaddress,nofootinbib,showpacs,longbibliography,10pt]{revtex4-2}

\usepackage[T1]{fontenc}
\usepackage[utf8]{inputenc}
\usepackage{lmodern}
\usepackage{physics}
\usepackage{verbatim}

\usepackage[dvipsnames]{xcolor}
\definecolor{linkcolor}{rgb}{0.0,0.3,0.5}
\usepackage[hypertexnames=false, unicode, colorlinks=true, linkcolor=linkcolor,
citecolor=linkcolor, filecolor=linkcolor,urlcolor=linkcolor,
pdfusetitle]{hyperref}
\usepackage{pgfplots}
\usepackage[all]{hypcap}
\usepackage{graphicx}
\usepackage{xspace}
\usepackage{fontawesome}
\usepackage[normalem]{ulem} 

\usepackage{natbib}

\usepackage{mathrsfs}
\usepackage{microtype}

\usepackage[english]{babel}

\usepackage{float}
\usepackage{orcidlink}

\usepackage{mathtools, bm}
\usepackage{booktabs}
\usepackage{tabularx}
\usepackage{ragged2e}
\usepackage{subcaption}
\usepackage{caption}
\DeclareCaptionJustification{justified}{\justifying}

\usepackage{xspace}

\begin{document}
	
	\preprint{}
	
	\title{Can Eccentric Binary Black Hole Signals Mimic Gravitational-Wave Microlensing?}
	
	\author{Anuj Mishra\orcidlink{0000-0002-2580-2339}}%
	\email{anuj.mishra@icts.res.in}
	\affiliation{%
		International Centre for Theoretical Sciences, Tata Institute of Fundamental Research, Bangalore 560089, India
	}%
	\affiliation{%
		The Inter-University Centre for Astronomy and Astrophysics (IUCAA), Post Bag 4, Ganeshkhind, Pune 411007, India
	}%
	\author{Apratim Ganguly\orcidlink{0000-0001-7394-0755}}%
	\email{apratim@iucaa.in}
	\affiliation{%
		The Inter-University Centre for Astronomy and Astrophysics (IUCAA), Post Bag 4, Ganeshkhind, Pune 411007, India
	}%
	
	
	\begin{abstract}
		Gravitational lensing in the wave-optics regime imprints characteristic frequency-dependent amplitude and phase modulations on gravitational-wave (GW) signals, yet to be detected by ground-based interferometers. Similar modulations may also arise from orbital eccentricity, raising the possibility of degeneracies that could lead to false microlensing claims. We investigate the extent to which eccentric binary black hole (BBH) signals can mimic microlensing signatures produced by an isolated point-mass lens. With a simulated population of eccentric signals using numerical relativity simulations and \texttt{TEOBResumS-Dal\'i} waveform model, we perform a Bayesian model-comparison study, supported by a complementary \textit{mismatch} analysis. We find a strong degeneracy for high eccentricities, low total masses, and high signal-to-noise ratios (SNRs): under these conditions, quasicircular microlensed model can be strongly favored over quasicircular unlensed model, even when the true signal is unlensed. For moderate SNRs ($\sim 30$), binaries with $M_\mathrm{tot}\lesssim 100\,M_\odot$ and eccentricity $e \gtrsim 0.4$ are particularly susceptible to misclassifications. In such cases, inferred microlens parameters exhibit well-constrained posteriors despite being unphysical. Crucially, the degeneracy is completely removed when the recovery uses waveform models that incorporate eccentricity, which overwhelmingly favors the eccentric hypothesis over microlensing. Our results demonstrate that any event exhibiting strong Bayesian evidence for microlensing should also be analyzed with eccentric waveform models and vice-versa to avoid false positives and biased astrophysical inference. This work contributes to developing robust strategies for interpreting signals in the era of precision GW astronomy.
	\end{abstract}
	
	\maketitle
	
	\section{\label{sec:intro}Introduction}
	The growing catalog of gravitational wave (GW) detections from compact binary coalescences (CBCs) by the LIGO~\cite{LIGOScientific:2014pky}, Virgo~\cite{VIRGO:2014yos}, and KAGRA~\cite{Somiya:2011np, Aso:2013eba, KAGRA:2018plz, KAGRA:2020tym} (LVK) Collaboration has opened a precision era of GW astronomy. With over $200$ confident detections reported to date \cite{LIGOScientific:2018mvr,  LIGOScientific:2020ibl, LIGOScientific:2021usb,  KAGRA:2021vkt, LIGOScientific:2025slb}, and with the anticipated sensitivities of next-generation detectors such as the Einstein Telescope (ET)~\cite{Punturo:2010zz, Hild:2010id}, Cosmic Explorer (CE)~\cite{Reitze:2019iox, LIGOScientific:2016wof, Regimbau:2016ike}, LISA~\cite{LISA:2017}, and DECIGO~\cite{Kawamura:2006up}, the detection of subtle physical signatures in the waveform is becoming increasingly important.
	Two such signatures of particular astrophysical interest are orbital eccentricity and gravitational lensing. 
	
	Gravitational lensing is a phenomenon that arises from the interaction between mass inhomogeneities along the line of sight and the propagation of radiation~\cite{Einstein:1936llh,Zwicky:1937zzb}. While gravitational lensing has been extensively observed in the electromagnetic spectrum~\cite[e.g.,][]{Walsh:1979nx,1988A&A...191L..19S}, it has yet to be confidently observed in the context of GWs~\cite{Hannuksela:2019kle,LIGOScientific:2021izm,LIGOScientific:2023bwz}, though several intriguing candidates have been proposed~\cite{Dai:2020tpj,Liu:2020par,Janquart:2023mvf}. The detection of lensing signatures in GWs holds immense significance: enable tests of general relativity~\cite{Mukherjee:2019wcg,Goyal:2020bkm}, probe fundamental physics~\cite{Baker:2016reh}, improve measurements of cosmological parameters~\cite{Liao:2017ioi,Jana:2022shb}, estimates of high-redshift merger rates~\cite{Mukherjee:2021qam}, and enhanced source localization~\cite{Hannuksela:2020xor}.

	In the geometric-optics regime, gravitational lensing by galaxy or galaxy cluster-scale lenses can produce multiple copies (or, \emph{images}) of a GW signal, known as \emph{strong} lensing, with distinct arrival times, magnifications, and phases~\cite{1971NCimB...6..225L, Ohanian:1974ys, Bernardeau:1999mh, Takahashi:2016jom, Liao:2017ioi, Dai:2017huk, Smith:2017jdz, Haris:2018vmn, Oguri:2018muv, Li:2018prc, Broadhurst:2018saj, Broadhurst:2019ijv, Broadhurst:2020cvm, Dai:2020tpj, Ezquiaga:2020gdt, More:2021kpb, Caliskan:2022wbh, Barsode:2024zwv}.
	These images fall into three categories: Type~I (minima), Type~II (saddle-point) \cite{Dai:2017huk, Dai:2020tpj, Ezquiaga:2020gdt, Vijaykumar:2022dlp}, and Type~III (maxima), whose phases differ by a Morse phase of $0$, $-\pi/2$, and $-\pi$, respectively.
	For smaller lens masses, where the GW wavelength becomes comparable to the Schwarzschild scale, wave-optics effects become important and produce frequency-dependent modulations in the observed signal, known as \emph{microlensing}~\cite{1986ApJ...307...30D, Nakamura:1997sw, Nakamura:1999uwi, 2003ApJ...595.1039T, Jung:2017flg, Shan:2020esq, Mishra:2021xzz, Meena:2022unp, Bondarescu:2022srx, Mishra:2023ddt, Mishra:2023vzo, Cremonese:2021puh, Cremonese:2021ahz,  Shan:2023ngi, Rao:2025poe}.
	Ground-based detectors are sensitive to microlensing by compact objects with masses $\sim 10$–$10^5~{\rm M}_\odot$, encompassing stellar-mass compact objects and intermediate-mass black holes. Searches in this band are especially valuable for constraining the compact dark matter fraction in this mass range~\cite{Basak:2021ten}, a region that is not yet well constrained.\footnote{We note that GWs also yield indirect constraints on compact dark matter fractions through the observed merger-rate density.}
	
	At the same time, several LVK events show hints of measurable eccentricity \cite{Lenon:2020oza, Romero-Shaw:2022xko, Iglesias:2022xfc, Dhurkunde:2023qoe, Gupte:2024jfe, Kacanja:2025kpr, McMillin:2025hof, LIGOScientific:2025brd, Tiwari:2025fua, Romero-Shaw:2025vbc}, even though the majority of binaries formed through isolated stellar evolution are expected to circularize by $\sim 10$~Hz~\cite{Peters:1964zz, Bethe:1998bn, LIGOScientific:2019dag}. While negligible eccentricities are expected for binaries formed through isolated stellar evolution~\cite{Mapelli:2021taw}, several alternative formation channels can yield highly eccentric systems ($e \gtrsim 0.5$ at 10 Hz) at small separations~\cite{Zevin:2018kzq}. Such channels include primordial black hole binaries~\cite{Cholis:2016kqi}, dynamical encounters in dense stellar environments~\cite{Wen:2002km}, and the evolution of isolated triple systems~\cite{Antonini:2013tea}. 
	
	While the rate of expected microlensed signals remains uncertain, the impact of non-zero orbital eccentricity is expected to become increasingly significant as we probe earlier stages of binary evolution or as detector sensitivities improve~\cite{Ng:2020qpk, Wang:2023tle, Yang:2024vfy}. Consequently, eccentric GW signals are expected to be detected by future ground- and space-based observatories such as CE, ET, DECIGO, and LISA. In case of eccentric signals, the usual oscillations in the strain at twice the orbital frequency are modulated by an oscillating envelope with a frequency lower than the orbital frequency, corresponding to the pericenter precession~\cite{Hinder:2017sxy}. Unlike the effects of binary masses, aligned spins, and tidal deformabilities, which are distinguishable from diffraction induced modulations because those do not cause oscillations in the amplitude and in the unwrapped phase of the frequency-domain waveform, the oscillation caused by the orbital eccentricity can potentially be mimicked by the modulations produced by wave-optics microlensing~\cite{Dai:2018enj}. If eccentricity is not modeled during inference, these effects may be confused and potentially lead to false microlensing claims.
	
	In this work, we focus on microlensing by an isolated point-mass lens and show that such biases can indeed arise when eccentricity is ignored in the analyses. Using Numerical Relativity and \texttt{TEOBResumS-Dal\'i} eccentric waveforms, together with both Bayesian model-comparison studies and \emph{mismatch} analyses, we examine whether unlensed eccentric BBH signals can be misidentified as microlensed when analyzed with quasi-circular templates, quantify the regions of parameter space where this confusion is most severe, and demonstrate that the degeneracy is fully broken when eccentric waveform models are used in recovery.
	
	The paper is organized as follows.  
	Section~\ref{sec:methodology} describes the microlensing framework and the computational setup. 
	In Sec.~\ref{sec:res} and Appendix~\ref{subsec:Ecc_pop}, we present the results of our Bayesian model-comparison study and mismatch analyses. We conclude in Sec.~\ref{sec:conclusion} with implications for future microlensing searches in the precision-GW era.

	\section{\label{sec:methodology}Methodology: Computational Setup and Bayesian Model comparison}
	
	In this section we summarize the microlensing formalism used to generate and recover microlensed waveforms, describe the injection and parameter estimation configuration employed throughout this work, and outline the Bayesian model comparison framework used to discriminate competing hypotheses.
	
	\subsection{\label{subsec:microlensing}Microlensing}
	In the wave–optics regime, microlensing modifies the frequency-domain GW strain by a complex frequency-dependent amplification factor. 
	The lensed waveform is related to the unlensed waveform by the multiplicative kernel
	\begin{equation}
		h_{\mathrm{ML}}(f;\pmb{\lambda},\pmb{\lambda}_{\mathrm{L}})
		= F(f;\pmb{\lambda}_{\mathrm{L}})\,h_{\mathrm{UL}}(f;\pmb{\lambda}),
		\label{eq:hml_definition_final}
	\end{equation}
	where $h_{\mathrm{UL}}$ and $h_{\mathrm{ML}}$ denote the unlensed and microlesned BBH waveform, respectively, $\pmb{\lambda}$ is the set of intrinsic and extrinsic binary parameters, and $\pmb{\lambda}_{\mathrm{L}}$ are the microlens parameters. 
	The complex amplification factor $F(f)$ encodes the frequency-dpendent diffraction and interference effects~\cite{1986ApJ...307...30D,Nakamura:1997sw,Nakamura:1999uwi,2003ApJ...595.1039T}.

	For concreteness, we adopt the canonical isolated point–mass lens model~\cite{2003ApJ...595.1039T}, in which the amplification factor is fully specified by two parameters: the redshifted lens mass $M_{\mathrm{L}}^{z} \equiv M_{\mathrm{L}}(1+z_{\mathrm{L}})$ and the dimensionless impact parameter $y$ (in units of the Einstein radius). In closed form, $F(f)$ is given by~\cite{2003ApJ...595.1039T}
	\begin{equation}
		\begin{aligned}
			F(f; M^{\rm z}_{\rm L}, y) &\equiv F\left(\omega,y\right)\\ &= \exp\bigg\{\frac{\pi \omega}{4} + 
			\frac{i\omega}{2}\left[\ln\left(\frac{\omega}{2}\right) - 
			2\phi_{\rm m}\left(y\right)\right]\bigg\} \\
			&\times\ \Gamma\left(1-\frac{i\omega}{2}\right)
			{}_{1}{F}_{1}\left(\frac{i\omega}{2},1;\frac{i\omega y^2}{2}\right),
		\end{aligned}
		\label{eq:point_mass_amp_final}
	\end{equation}
	where $\omega = 8\pi GM^{\rm z}_{\rm L} f/c^3$ represents the dimensionless frequency that depends solely on  $M^{\rm z}_{\rm L}$ for a given dimensionful frequency $f$, and $\phi_{\rm m}(y)$ is a frequency-independent quantity depending only on $y$, given by $\phi_{\rm m}(y) = (x_{\rm m} - y)^2/2 - \ln(x_{\rm m})$, where $x_{\rm m} = \left(y+\sqrt{y^2+4}\right)/2$.
	Equations~\eqref{eq:hml_definition_final}-\eqref{eq:point_mass_amp_final} are sufficient to generate microlensed strains from any frequency-domain unlensed waveform model.
	
	All microlensed injections and parameter inference studies in this work use a custom frequency-domain implementation that augments standard BBH models with the two microlens parameters $(M_{\mathrm{L}}^{z},y)$. The implementation is publicly available in the \texttt{GWMAT} package.\footnote{\url{https://git.ligo.org/anuj.mishra/gwmat/}}
	
	Unless otherwise stated, we adopt weakly informative priors commonly used in microlensing searches:
	\begin{equation}
		\begin{aligned}
			p(\log_{10}M_{\mathrm{L}}^{z}) &\propto \mathrm{Uniform}(-1,5),\\
			p(y) &\propto y\quad\text{for}\quad y\in(10^{-3},5).
		\end{aligned}
		\label{eq:lensing_priors_final}
	\end{equation}
	The $p(y)\propto y$ prior follows from simple geometric/isotropic arguments for random source–lens alignments\cite{Lai:2018rto}. For cases where we observe the corresponding 1D-marginalized posteriors \emph{railing} against the prior bounds, we broaden the priors accordingly to avoid artificial truncation.

	
	
	\subsection{\label{subsec:injections_and_pe}Injection strategy and parameter–estimation setup}
	To study the biases in microlensing searches ignoring eccentricity in the recovery templates, we consider simulated BBH signals, or \textit{injections}. These injections are done in the detector network of LIGO-Virgo with the projected Advanced LIGO and Virgo sensitivities\footnote{For LIGO detectors, we use the PSD given in \href{https://dcc.ligo.org/public/0165/T2000012/002/aligo_O4high.txt}{here}. While for Virgo, we use the PSD available \href{https://dcc.ligo.org/public/0165/T2000012/002/avirgo_O4high_NEW.txt}{here}.}~\cite{Abbott:2020search}. 
	To isolate modeling biases and avoid noise systematics, i.e., biases due to specific noise realizations, we do not add noise to our injections. In other words, we assume a realization of Gaussian noise that takes its mean value (zero) at every time step. This avoids stochastic fluctuations due to particular noise realizations and highlights systematic waveform-modeling effects.
	
	Throughout this work, we use the publicly available Bayesian inference library \texttt{Bilby}~\cite{Ashton:2018jfp,Smith:2019ucc} for performing parameter estimation runs and computing the evidences. 
	Specifically, we use \texttt{Dynesty}~\cite{2020MNRAS.493.3132S} nested sampler with the `acceptance-walk' method for the Markov-Chain Monto-Carlo (MCMC) evolution as implemented in \texttt{Bilby}. 
	We evaluate the likelihood from a lower frequency of $f_\mathrm{low}=20~\rm{Hz}$ up to the Nyquist frequency of $f_\mathrm{high}=1024~\rm{Hz}$, under the assumption of stationary and Gaussian noise which is uncorrelated across detector.
	Our default priors are chosen to be agnostic and sufficiently wide to cover the region of the parameter space where the posteriors have support. For quasicircular BBH waveform models, which are usually $15$ dimensional, they are uniform in (redshifted) component masses, uniform in spin magnitudes, isotropic in spin orientations, isotropic in binary orientation, uniform in merger time and coalescence phase, isotropic in sky location, and our distance prior corresponds to a uniform merger rate in comoving volume and time.
	
	\subsection{\label{subsec:model_selection}Bayesian model comparison and interpretation}
	To compare between any two models/hypotheses, say, $\mathcal{H}_{1}$ and $\mathcal{H}_{2}$, we utilize Bayesian model comparison scheme.
	Specifically, we make use of the Bayesian evidence (or marginal likelihood) $\mathcal{Z}=\int d\pmb{\theta}\pi(\pmb{\theta})\mathcal{L}(\pmb{d} | \pmb{\theta})$, which is an $n-$dimensional integral. Here, $\pmb{\theta}\in \mathbb{R}^n$ is the parameter space describing the signal, $\pi(\pmb{\theta})$ is the prior expectation we have about $\pmb{\theta}$, and $\mathcal{L}(\pmb{d}|\pmb{\theta})$ is the likelihood of observing the data $\pmb{d}$ given $\pmb{\theta}$.
	Under the assumption that the two models are equally likely, we set their prior probabilities to be equal (i.e., $p(\mathcal{H}_\mathrm{1})=p(\mathcal{H}_\mathrm{2})$). The two models can then be compared using their Bayes factors as:
	\begin{align}
		\frac{p(\mathcal{H}_2|\pmb{d})}{p(\mathcal{H}_\mathrm{1}|\pmb{d})} &= \frac{p(\pmb{d} | \mathcal{H}_2)}{p(\pmb{d} | \mathcal{H}_\mathrm{1})} = \frac{\mathcal{Z}_\mathrm{\mathcal{H}_2}}{\mathcal{Z}_\mathrm{\mathcal{H}_1}} \equiv \mathcal{B}^{\mathcal{H}_2}_{\mathcal{H}_1},\\
		\Rightarrow \hspace{0.5cm} \log_{10}\mathcal{B}^{\mathcal{H}_2}_{\mathcal{H}_1}  &= \log_{10}\mathcal{Z}_\mathrm{\mathcal{H}_2} -\log_{10}\mathcal{Z}_\mathrm{\mathcal{H}_1}.
	\end{align}
	In practice we compare (i) the quasicircular microlensed (QCML) hypothesis $\mathcal{H}_{\rm QCML}$ to the quasicircular unlensed (QCUL) hypothesis $\mathcal{H}_{\rm QCUL}$, and (ii) the eccentric unlensed (EccUL) hypothesis $\mathcal{H}_{\rm EccUL}$ to $\mathcal{H}_{\rm QCUL}$, thereby quantifying whether eccentricity can masquerade as microlensing under standard quasicircular analyses.
	
	\paragraph*{Interpretation of Bayes Factors:}
	In this work, we mainly use Jeffreys’ criterion~\cite{Jeffreys:1939xee} for interpreting Bayes factors. However, it only provides a conservative estimate for two main reasons:
	\begin{itemize}
		\item Since the injections are performed in zero noise, fluctuations associated with specific noise realizations are not accounted for. Such fluctuations can increase the uncertainty in the Bayes factor and may drive it to even higher values.
		\item Evidence integrals depend on priors; for $\mathcal{H}_{\rm QCML}$, adding microlens parameters enlarges the prior volume and introduces an Occam penalty. Moreover, The $p(y)\propto y$ prior preferentially weights weakly microlensed (large $y$) regions where $F(f)\to1$, reducing evidence for microlensing unless the data show substantial likelihood gain at small $y$.
	\end{itemize}
	Therefore, we expect the preference for the microlensing hypothesis to increase when an agnostic uniform prior on $y$ is used, and the uncertainty to increase in the presence of noise. The Bayes factors reported here should thus be regarded as conservative estimates.

	\begin{table*}
		\caption{\label{tab:NR_injs_res}
			Bayesian model comparison between the QCUL and QCML hypotheses for six zero-noise non-spinning numerical-relativity injections from the SXS catalog, corresponding to QCUL and EccUL signals, each with mass ratios $(q)$ of $\{1,2,3\}$.
			The total binary masses $(M)$ for the eccentric injections are chosen such that all the signal content above $20~\mathrm{Hz}$ is retained~(see Sec.~2G of \cite{Shaikh:2023ypz} for further details). We set the same $M$ for both EccUL and the corresponding QCUL injection with the same $q$ for comparison.
			For each eccentric injection, we quote the gauge-independent eccentricity measured at $20~\mathrm{Hz}$ $(e^{\mathrm{gw}}_{20~\mathrm{Hz}})$ using the \texttt{gw\_eccentricity} package. The Bayes factor comparing the QCML and QCUL hypotheses is denoted as $\log_{10}\mathcal{B}^\mathrm{QCML}_\mathrm{QCUL}$.
			Additionally, for each case, we list the inferred microlens parameters, $(\log_{10}M^{\mathrm{z}}_{\mathrm{L}})$ and $(y)$, along with the Jensen–Shannon divergence (JSD) values between their 1D marginalized posteriors and corresponding priors. Finally, we also provide the Bayes factor values between the EccUL and QCUL models for reference. The extrinsic parameters of the signal are kept same as that of \texttt{GW150914}, except that of the luminosity distance, which is scaled to achieve a network optimal SNR of $100$, and the inclination is set to $0$.
		}
		\begin{ruledtabular}
			\begin{tabular}{c c c c c c c c c c c}
				SXS ID & Orbit type & $q$ & $M/$M$_\odot$ &$e^{\rm gw}_{20{\rm Hz}}$ & $\log_{10}\mathcal{B}^\mathrm{QCML}_\mathrm{QCUL}$ & $\log_{10}M^{\rm z}_{\rm L}$ & JSD($M^{\rm z}_{\rm L}$)\footnotemark[1] & $y$ & JSD($y$)\footnotemark[1] & $\log_{10}\mathcal{B}^\mathrm{EccUL}_\mathrm{QCUL}$ \\
				\hline\\[-0.25cm] 
				$\texttt{SXS:BBH:1132}$ & quasicircular & $1$ & $75.0$ & - & $-0.55$ & $-0.16_{-0.57}^{+0.60}$ & $0.57$ & $3.60_{-1.46}^{+0.99}$ & $0.05$ & $-0.38$ \\[5pt]
				
				$\texttt{SXS:BBH:1222}$ & quasicircular & $2$ & $77.6$ & - & $-0.55$ & $-0.13_{-0.59}^{+0.67}$ & $0.57$ & $3.63_{-1.43}^{+0.97}$ & $0.05$ & $-0.59$ \\[5pt]
				
				$\texttt{SXS:BBH:2265}$ & quasicircular & $3$ & $87.1$ & - & $-0.50$ & $-0.12_{-0.60}^{+0.75}$ & $0.55$ & $3.68_{-1.46}^{+0.94}$ & $0.05$ & $-0.34$ \\[5pt]
				
				$\texttt{SXS:BBH:2524}$ & Eccentric & $1$ & $75.0$ & $0.2893$ & $13.86$ & $4.88_{-0.02}^{+0.02}$ & $0.79$ & $2.64_{-0.09}^{+0.12}$ & $0.72$ & $471.81$ \\[5pt]
				
				$\texttt{SXS:BBH:3957}$ & Eccentric & $2$ & $77.6$ & $0.2912$ & $8.91$ & $1.89_{-0.04}^{+0.04}$ & $0.79$ & $2.58_{-0.16}^{+0.18}$ & $0.67$ & $488.81$ \\[5pt]
				
				$\texttt{SXS:BBH:2544}$ & Eccentric & $3$ & $87.1$ & $0.2914$ & $32.32$ & $5.00_{-0.01}^{+0.01}$ & $0.79$ & $2.14_{-0.04}^{+0.05}$ & $0.77$ & $544.51$ \\
				
			\end{tabular}
		\end{ruledtabular}
		\footnotetext[1]{Jensen–Shannon divergence value between the prior and the 1D marginalized posterior density for the parameter quoted in parenthesis.}
	\end{table*}
	
	\section{\label{sec:res}Results}
	We present two complementary sets of studies designed to quantify the extent to which orbital eccentricity can mimic wave–optics gravitational lensing and to determine when that degeneracy can be broken. In Sec.~\ref{subsec:NR_injs}, we analyze highly reliable NR eccentric injections, while in Sec.~\ref{subsec:ecc_TEOBResumS_injs}, we perform a larger controlled injection campaign using \texttt{TEOBResumS-Dal\'i} waveforms to explore the degeneracy across total mass, mass ratio, and eccentricity. In both cases, we first analyze the injections with quasicircular templates and then with eccentric templates to demonstrate how including the correct physics removes the confusion.
	
	\subsection{\label{subsec:NR_injs}Numerical Relativity Injections}
	In this section, we examine whether a standard (unlensed) eccentric BBH GW signal can be incorrectly classified as a quasicircular microlensed signal when eccentricity is ignored during the analysis.
	To perform this study, we use some of the most accurate eccentric GW signals currently available, obtained entirely from numerical relativity (NR) simulations. Specifically, we select three non-spinning eccentric simulations from the latest (third) SXS catalog release~\cite{Scheel:2025jct}, with SXS IDs \{\texttt{SXS:BBH:2524}, \texttt{SXS:BBH:3957}, \texttt{SXS:BBH:2544}\}, corresponding to mass ratios $q=\{1,2,3\}$, respectively, and matching quasicircular counterparts \{\texttt{SXS:BBH:1132}, \texttt{SXS:BBH:1222}, \texttt{SXS:BBH:2265}\}. 
	For the injections, we utilize the highest available resolution of these simulations and use extrapolated waveforms with extrapolation order N=2, considering only the dominant $(\ell,|m|)=(2,2)$ mode.
	
	For each eccentric NR case, the total redshifted binary mass is chosen such that the periastron frequency of the dominant $(\ell,|m|)=(2,2)$ mode crosses $20\,$Hz at the NR relaxation time\footnote{The time removed from the start of the NR time series to ensure the junk radiation is negligible~\cite{Scheel:2025jct}.} giving $M_{\rm tot}\approx\{75.0,77.6,87.1\}~\mathrm{M}_\odot$ (we follow the prescription in Sec.~2G of~\cite{Shaikh:2023ypz}). The corresponding gauge independent eccentricities measured at $20~$Hz, computed using the \texttt{gw\_eccentricity} package~\cite{Shaikh:2023ypz, Shaikh:2025tae}, are $e^{\mathrm{gw}}_{20\rm Hz}\approx\{0.2893,0.2912,0.2914\}$. 
	
	We choose the inclination to be $0$ (face-on binary) so that only the $m=2$ [spin-$(-2)$-weighted] spherical harmonic modes of the signal contribute, avoiding higher-order modes that would require significantly longer NR simulations for the masses considered here.
	The luminosity distance is adjusted to produce a network optimal SNR of $100$ for each injection. Other extrinsic parameters are set to the median posterior values inferred for \texttt{GW150914}\footnote{Posterior samples available at \href{https://zenodo.org/records/6513631}{Zenodo}. We use the median of the 1D-marginalized posteriors from the `C01:Mixed' channel of `IGWN-GWTC2p1-v2-GW150914\_095045\_PEDataRelease\_mixed\_cosmo.h5'.}~\cite{LIGOScientific:2016aoc}.
	
	For parameter estimation and evidence computation with \texttt{Bilby}, we use only the dominant $(\ell, |m|)=(2, 2)$ mode (same as in the injections). To compare $\mathcal{H}_{\rm QCML}$ and $\mathcal{H}_{\rm QCUL}$, we compute $\log_{10}\mathcal{B}^{\rm QCML}_{\rm QCUL}$, employing the \texttt{IMRPhenomXP} waveform model~\cite{Khan:2018fmp, Garcia-Quiros:2020qpx, Pratten:2020ceb}. The sampler settings are \texttt{nlive}=1000, \texttt{n-accept}=60, and \texttt{n-parallel}=3 per injection, with a stopping criterion of $\Delta \log \mathcal{Z}<0.1$. 
	
\begin{figure*}[t]
		\centering
		\begin{subfigure}[t]{0.49\textwidth}
			\centering
			\includegraphics[width=\linewidth]{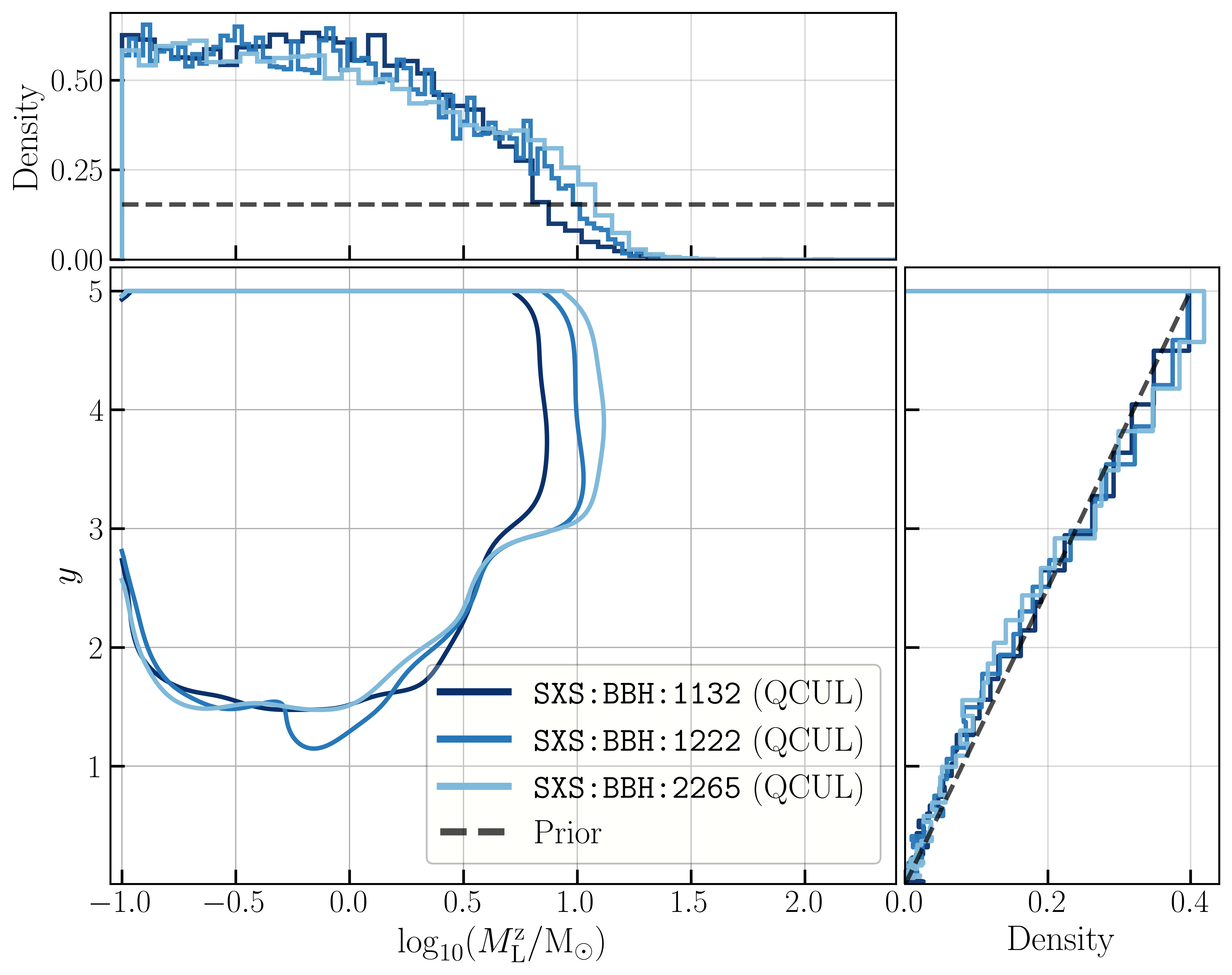}
		\end{subfigure}
		\hfill
		\begin{subfigure}[t]{0.49\textwidth}
			\centering
			\includegraphics[width=\linewidth]{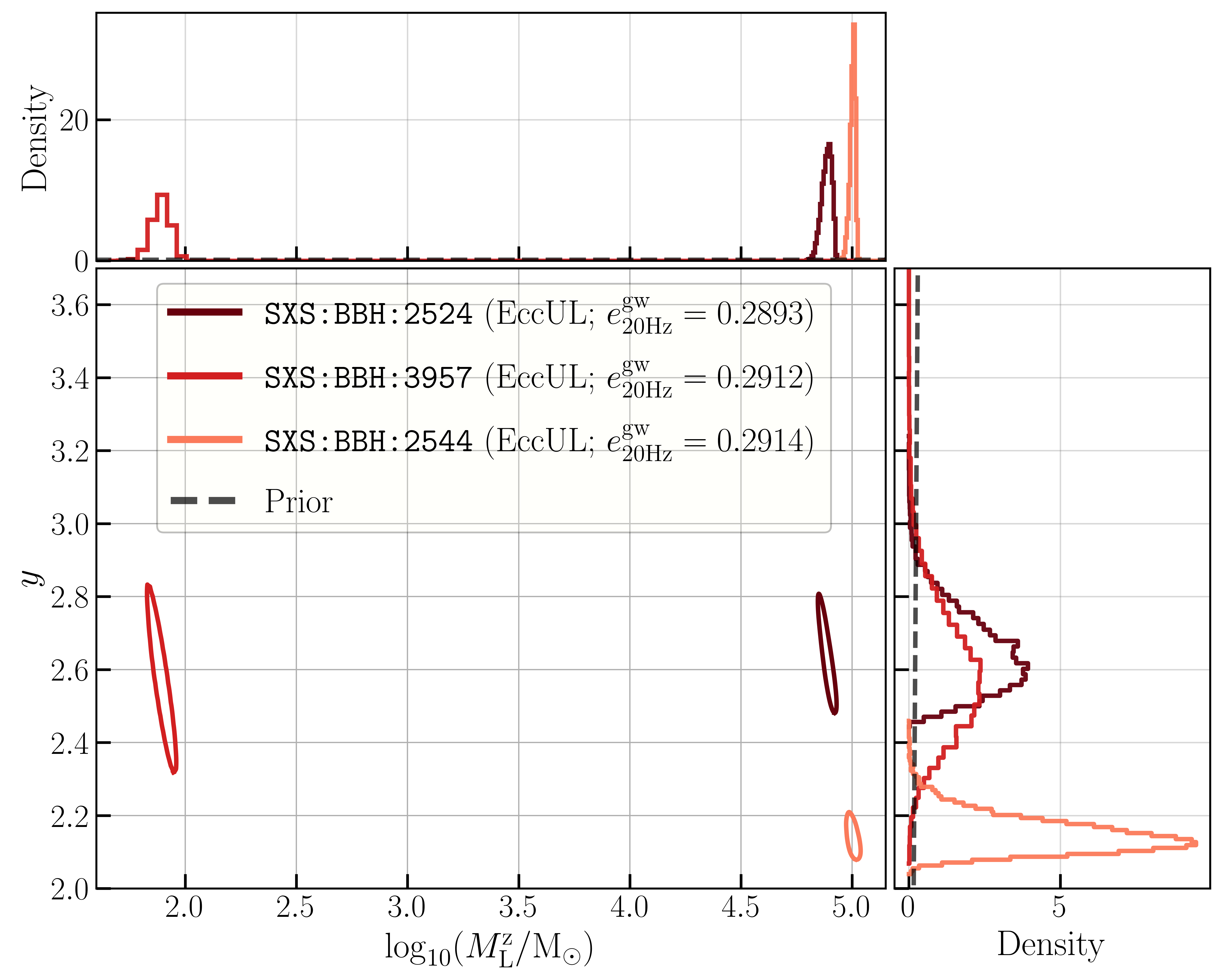}
		\end{subfigure}
		\caption{Inferred microlens parameters, $\log_{10}M^{\rm z}_{\rm L}$ and $y$, for three non-spinning QCUL NR injections (left panel) and three non-spinning EccUL NR injections (right panel), generated using SXS simulations (SXS IDs mentioned in the legend). The bottom-left subpanels display the inferred $1\sigma$ credible regions for the lens parameters, while the adjacent panels show the corresponding $1$D marginalized posteriors. For eccentric simulations, we additionally report the gauge-independent eccentricity measured at $20~\mathrm{Hz}$ $(e^{\mathrm{gw}}_{20~\mathrm{Hz}})$ using the \texttt{gw\_eccentricity} package.}
		\label{fig:MicL_params_NR_Injs_1}
	\end{figure*}

	
	Table~\ref{tab:NR_injs_res} summarizes the model comparison results. Notably, all eccentric NR injections show very strong support for the microlensing hypothesis $(\log_{10}\mathcal{B}^{\rm QCML}_{\rm QCUL}> 1)$. 
	The corresponding inferred microlens parameters (red shaded curves; right panel in Fig.~\ref{fig:MicL_params_NR_Injs_1}) also differ noticeably from their prior distributions (black dashed curves), with well-localized one-dimensional marginalized posteriors. This behavior is quantified by the Jensen–Shannon divergence (JSD) between the prior and posterior distributions. For $y$, JSD increases from $\sim0.05$ for the QCUL injections to $\gtrsim 0.7$ for the EccUL injections. For $\log_{10}M^{\rm z}_{\rm L}$, it increases from $\sim0.55$ (QCUL) to $\sim0.79$ (EccUL). The comparatively larger JSD for $\log_{10}M^{\rm z}_{\rm L}$ in the QCUL injections is expected, as this parameter is typically better constrained in the weak-microlensing regime (see Appendix A of~\cite{Mishra:2023ddt}). By contrast, all quasicircular injections yield $\log_{10}\mathcal{B}^\mathrm{QCML}_\mathrm{QCUL}<0$, decisively rejecting the microlens hypothesis, as expected. Their inferred 1D-marginalized posteriors for $y$ closely follow the prior (blue shaded curves; left panel in Fig.~\ref{fig:MicL_params_NR_Injs_1}), and their inferred $\log_{10}M^{\rm z}_{\rm L}$ values remain below unity, consistent with the expectations for unlensed systems.
	
	Lastly, we also perform an eccentric–versus–quasicircular model–comparison study by computing $\log_{10}\mathcal{B}^\mathrm{EccUL}_\mathrm{QCUL}$. To avoid waveform–model systematics, we use the same waveform model for both hypotheses: \texttt{TEOBResumS-Dal\'i}\footnote{Version v1.1.0, commit hash \texttt{6de7c26}.}~\cite{Damour:2014sva, Nagar:2015xqa, Nagar:2018zoe, Nagar:2019wds, Nagar:2020pcj, Riemenschneider:2021ppj, Nagar:2023zxh}, a state-of-the-art eccentric IMR waveform model that is currently publicly available.
	For the EccUL hypothesis, the eccentricity is treated as a free parameter with a uniform prior in $[0, 0.98]$, defined at the periastron frequency, which is set equal to the waveform starting frequency of $f_{\rm start}=20~$Hz; the true anomaly has a uniform prior on $[0, 2\pi]$; the in-plane spin components are fixed to zero; and the aligned spin components are allowed to vary in $[-0.9, 0.9]$ with a prior uniform in magnitude and isotropic in orientation (uniform in the cosine of the tilt angle). Rest of the priors are chosen to be agnostic and sufficiently wide around the true injected values.
	We use the same mode content to that used in the injections, and set the likelihood low-frequency cutoff $f_{\rm low} = 20$ Hz. The sampling frequency is set to $2048$ Hz.
	For the QCUL hypothesis, we adopt the same settings, except that both the eccentricity and anomaly parameters are fixed to zero.
	Since \texttt{TEOBResumS-Dal\'i} is relatively slower waveform to evaluate $(\mathcal{O}(10^{-1}~\rm{s})$ per waveform evaluation$)$, we use $\texttt{nlive}=500$ and relax the absolute and relative error tolerances of its ordinary differential equations (ODEs) integrator from their default values of $10^{-13}$ and $10^{-12}$ to $10^{-8}$ and $10^{-7}$~(see \cite{OShea:2021faf} for more details), respectively, reducing the evaluation cost to $\mathcal{O}(10^{-2})~$s per evaluation. 
	
	The resulting Bayes factors, listed in the final column of Table~\ref{tab:NR_injs_res}, show that all eccentric injections yield $\log_{10}\mathcal{B}^{\rm EccUL}_{\rm QCUL}\gg 1$, indicating strong evidence for eccentricity. Conversely, quasicircular injections correctly show no support for eccentricity $(\log_{10}\mathcal{B}^{\rm EccUL}_{\rm QCUL}< 0)$. Moreover, for all eccentric injections we find $\log_{10}\mathcal{B}^{\rm EccUL}_{\rm QCUL}\gg \log_{10}\mathcal{B}^{\rm QCML}_{\rm QCUL}$, demonstrating that eccentric models provide far better fits than microlensed models. 
	This indicates that the degeneracy between eccentricity-induced and microlensing-induced waveform features is \textit{weak} and can be broken when the correct waveform family is used in the analysis. 
	This conclusion is further corroborated by the fact that, at lower network SNRs (e.g., SNR $15$), we do not observe any significant support for the microlensing hypothesis $(\log_{10}\mathcal{B}^{\rm QCML}_{\rm QCUL}<0)$ when repeating the same exercise (although we do not explicitly present those results here).
	
	The findings in this section highlight the presence of degeneracies between the observational signatures of microlensing and eccentricity in GWs from a BBH system, underscoring the need for a more comprehensive investigation. We address this in the subsequent sections.
	
	\begin{figure*}[t]
		\centering
		\begin{subfigure}[t]{0.49\textwidth}
			\centering
			\includegraphics[width=\linewidth]{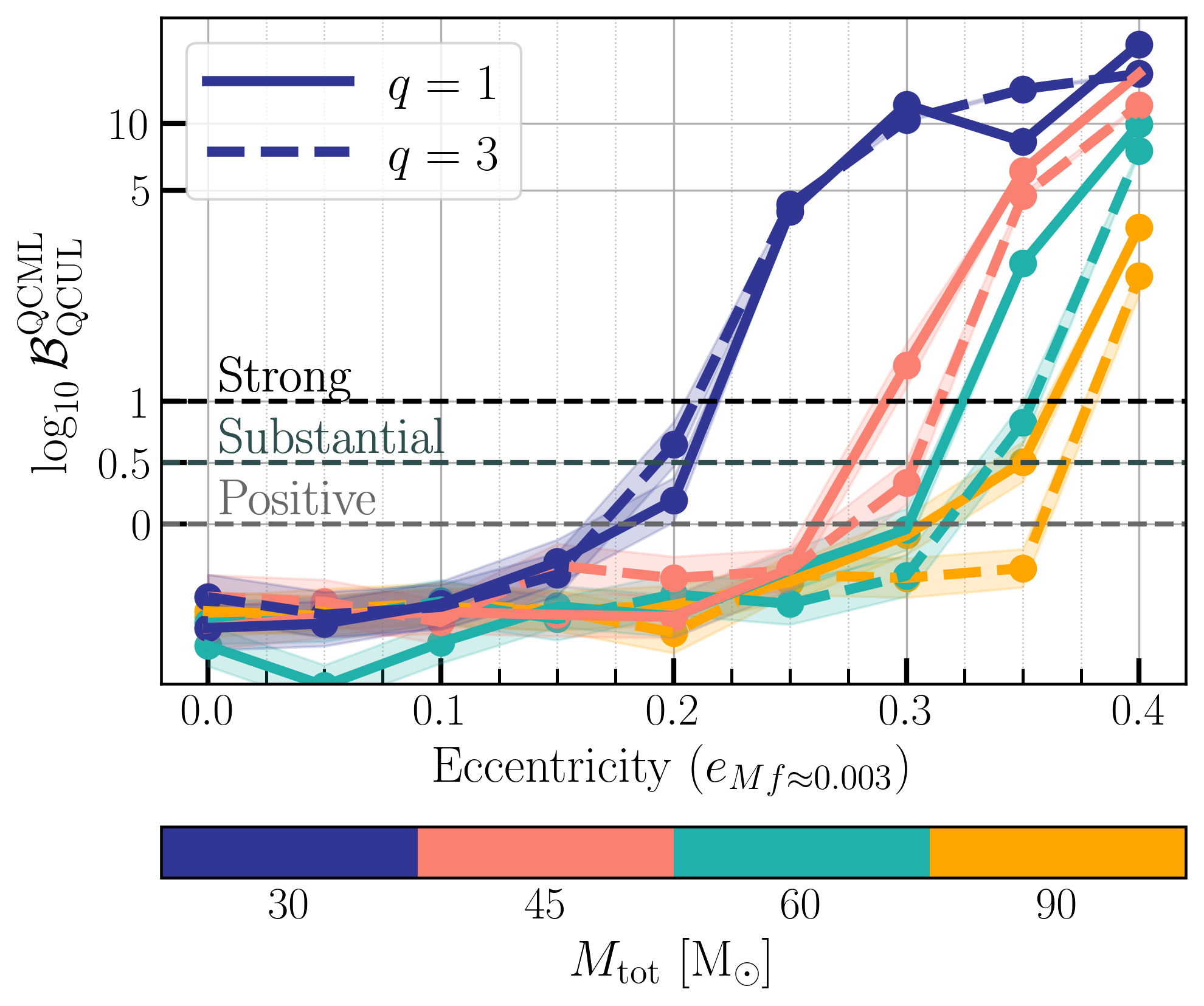}
			\caption{
				Bayesian model comparison between QCML and QCUL hypotheses performed using the \texttt{TEOBResumS-Dal\'i} waveform model, where in-plane spin components, eccentricity, and the true anomaly are fixed to zero during inference.
			}
			\label{fig:BF_QCML_QCUL_TEOB}
		\end{subfigure}
		\hfill
		\begin{subfigure}[t]{0.49\textwidth}
			\centering
			\includegraphics[width=\linewidth]{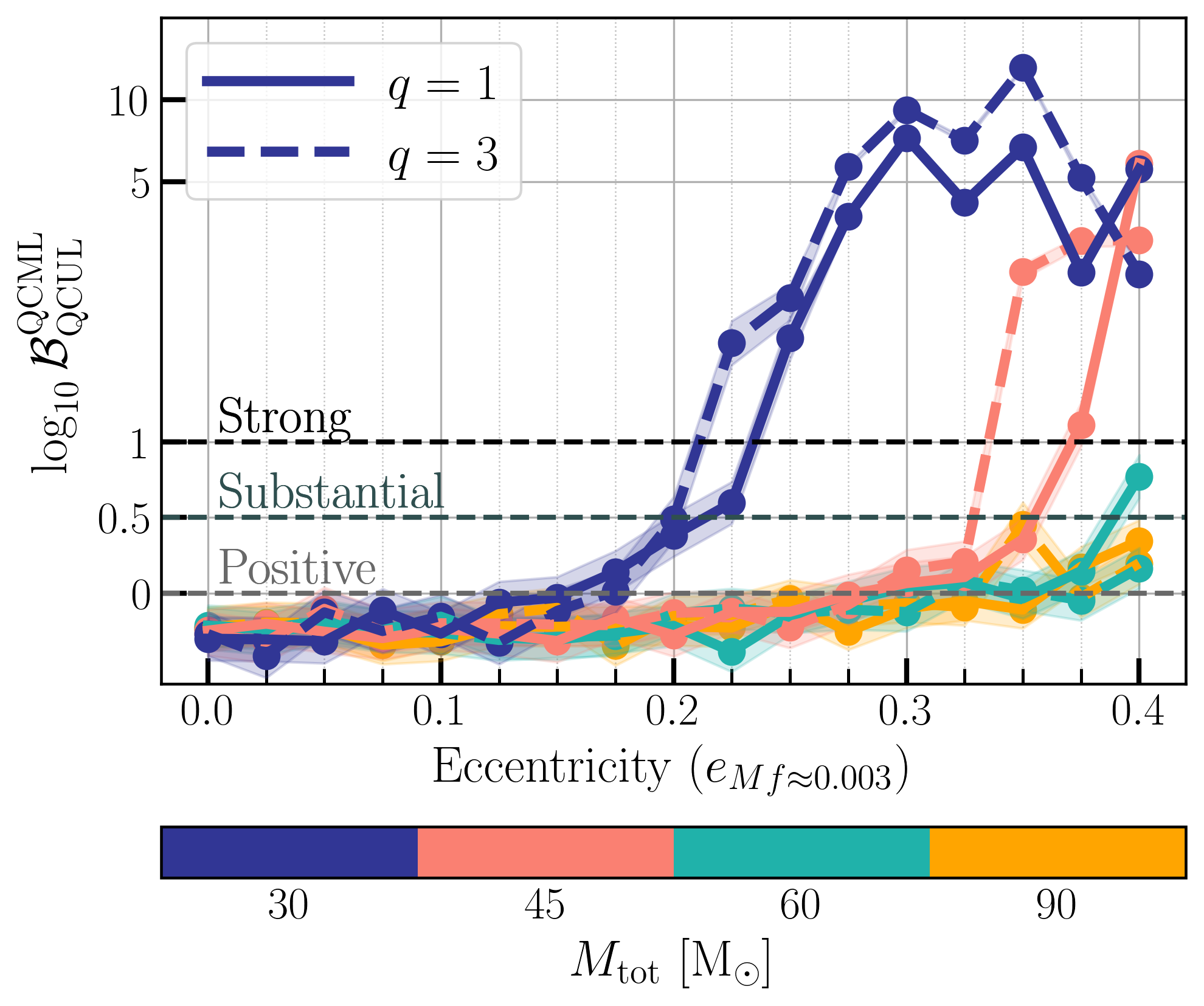}
			\caption{
				Bayesian model comparison between QCML and QCUL hypotheses using the \texttt{IMRPhenomXPHM-SpinTaylor} waveform approximant, where all 15 parameters are utilized during inference.
			}
			\label{fig:BF_QCML_QCUL_XPHM}
		\end{subfigure}
		
		\caption{
			Bias in microlensing searches arising from the presence of eccentricity in the signal. The variation in the Bayes factors between the QCML and QCUL hypotheses, $\log_{10}\mathcal{B}^\mathrm{QCML}_\mathrm{QCUL}$, is shown as a function of the eccentricity $(e)$ (x-axis), total binary mass $(M_\mathrm{tot})$ (color scale), and mass ratio $(q)$ (solid vs.~dashed lines). Eccentricity is defined at a dimensionless frequency of $\sim 0.003$ at apastron. 
			The uncertainty in the estimated $\log_{10}\mathcal{B}^\mathrm{QCML}_\mathrm{QCUL}$ is represented by the translucent bands around each curve.
			The injections are performed using the \texttt{TEOBResumS-Dal\'i} waveform model, while the recovery (evidence computation) is done using two different quasicircular models: (a) \texttt{TEOBResumS-Dal\'i} with eccentricity and related parameters fixed to zero, and (b) \texttt{IMRPhenomXPHM-SpinTaylor} allowing all 15 parameters to vary. 
			These results demonstrate that neglecting eccentricity in waveform modeling can lead to spurious support for the microlensing hypothesis, especially at higher eccentricities.
		}
		\label{fig:BF_QCML_QCUL_combined}
	\end{figure*}
	
	\begin{figure*}[t]
		\centering
		\begin{subfigure}[t]{0.99\textwidth}
			\centering
			\includegraphics[width=\linewidth]{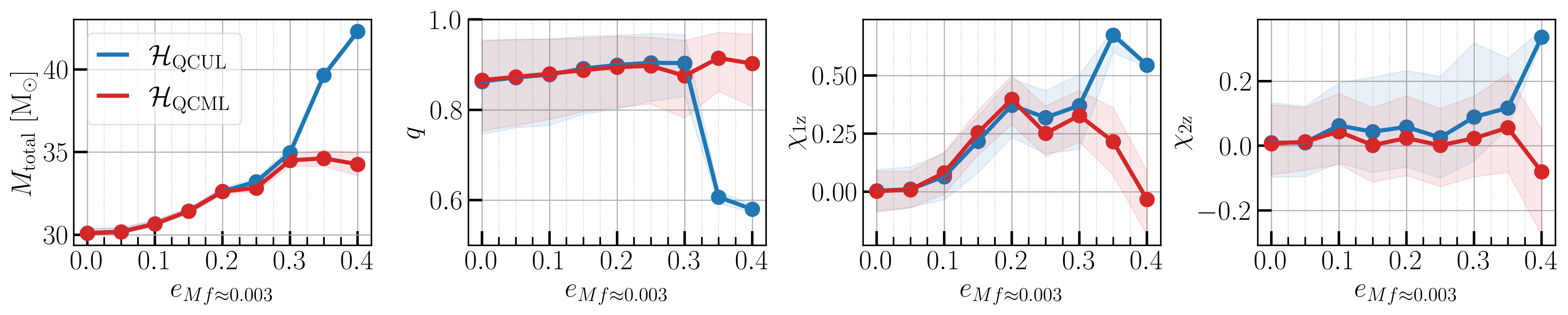}
			\caption{
				Analysis using \texttt{TEOBResumS-Dal\'i} waveform model, where in-plane spin components, eccentricity, and the true anomaly are fixed to zero during inference.
			}
			\label{fig:bias_TEOB}
		\end{subfigure}
		\hfill
		\begin{subfigure}[t]{0.99\textwidth}
			\centering
			\includegraphics[width=\linewidth]{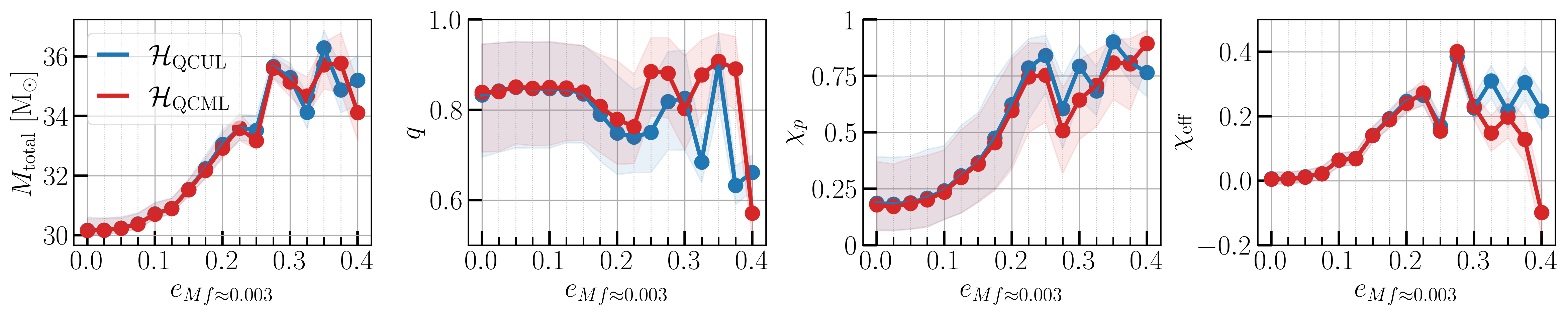}
			\caption{
				Analysis using \texttt{IMRPhenomXPHM-SpinTaylor} waveform approximant, where all 15 parameters are varied during inference.
			}
			\label{fig:bias_XPHM}
		\end{subfigure}
		
		\caption{
			Comparison of the inferred intrinsic parameters as a function of eccentricity ($e_{M f \approx 0.003}$) when eccentric injections are analyzed under the quasicircular microlensed (QCML; $\mathcal{H}_{\rm QCML}$) and quasicircular unlensed (QCUL; $\mathcal{H}_{\rm QCUL}$) hypotheses.
			For the \texttt{TEOBResumS-Dal\'i} waveform model [panel (a)], the intrinsic parameters shown are the total mass $(M_\mathrm{tot})$, mass ratio $(q)$, and aligned spin components $(\chi_{1z}, \chi_{2z})$.
			For the \texttt{IMRPhenomXPHM-SpinTaylor} model [panel (b)], we instead show the total mass $(M_\mathrm{tot})$, mass ratio $(q)$, effective precession spin parameter $(\chi_{\rm p})$, and effective aligned-spin parameter $(\chi_{\rm eff})$.
			All results correspond to non-spinning injections with $\{M_\mathrm{tot} = 30~{\rm M}_\odot, q = 1\}$.
			The points denote the median values of the one-dimensional marginalized posteriors, and the shaded bands indicate the corresponding $1\sigma$ uncertainties.
			In panel (a), in-plane spins, eccentricity, and true anomaly are fixed to zero during inference, whereas in panel (b), all 15 parameters are varied.
		}
		\label{fig:bias_combined}
	\end{figure*}
	
	
	\subsection{\label{subsec:ecc_TEOBResumS_injs}Injection Study using \texttt{TEOBResumS-Dal\'i}}
	To explore the degeneracy over a broader region of parameter space of eccentricity, binary mass and the mass-ratio, we utilize \texttt{TEOBResumS-Dal\'i}~\cite{Damour:2014sva, Nagar:2015xqa, Nagar:2018zoe, Nagar:2019wds, Nagar:2020pcj, Riemenschneider:2021ppj, Nagar:2023zxh} waveform model to perform eccentric injections
	
	
	For injections, we choose four distinct binary mass values $M_\mathrm{tot}$/M$_\odot\in \{30,45,60,90\}$, two mass ratio values $q \in \{1,3\}$, and eccentricity $(e)$ values 
	sampled as follows: for analyses using \texttt{IMRPhenomXPHM-SpinTaylor}, $e$ ranges from $0$ to $0.4$ in steps of $0.025$, while for \texttt{TEOBResumS-Dal\'i}, $e$ ranges from $0$ to $0.4$ in steps of $0.05$. This choice is made because, as mentioned in Sec.~\ref{subsec:NR_injs}, \texttt{TEOBResumS-Dal\'i} is typically slower to evaluate than \texttt{IMRPhenomXPHM-SpinTaylor}.
	These eccentricity values are defined at a dimensionless apastron frequency of $Mf_{\rm ref}\approx 0.003$ evaluated for the fundamental $(\ell=2,|m|=2)$ mode at apastron, which corresponds to $10~$Hz for a $60~$M$_\odot$ BBH system\footnote{We choose this value because according to Table I of \cite{Chiaramello:2020ehz}, \texttt{TEOBResumS-Dal\'i} has been shown to be faithful against NR simulations up to an eccentricity of $\sim0.3$ at $10~$Hz for a $60~$M$_\odot$ BBH system.}. This ensures that we define eccentricity at the same point in the evolution of a binary.
	We consider only $(\ell,|m|)=\{(2,2),(2,1),(3,2),(3,3),(4,4)\}$ modes for the injection.
	The luminosity distance is adjusted to produce a true network optimal SNR of $30$ for each injection. Other extrinsic parameters are set to the median posterior values inferred for \texttt{GW150914}~\cite{LIGOScientific:2016aoc}.
	We note that ideally one should utilize full signal content above $f_{\rm low}=20~$Hz, which implies considering the signal from the point when its periastron frequency of the fundamental mode crosses $f_{\rm start}=f_{\rm low}\cdot(2/m_{\rm max})=10~$Hz, where $m_{\rm max}$ corresponds to the largest $m$ among all included modes (see Eq.~18 in \cite{Shaikh:2023ypz}), as we did in the last section~\ref{subsec:NR_injs}.
	However, in this section, we consider the waveform from the point when the apastron frequency of the fundamental mode crosses $f_{\rm ref}$.\footnote{In \texttt{TEOBResumS-Dalí}, the reference frequency $f_{\rm ref}$ where the eccentricity is defined and the starting frequency $f_{\rm start}$ of the fundamental mode are same.}
	We do this mainly to decrease computational expense. If we had instead started the waveform at the point where the periastron frequency crosses $f_{\rm ref}$, then both the waveform duration and the waveform evaluation time would increase with eccentricity, making PE runs significantly costlier.
	
	\subsubsection{\label{subsubsec:QC_recs}Inference with quasicircular templates for Eccentric Injections}
	To isolate the effect of ignoring eccentricity in microlensing searches, we first analyze the signals using the quasicircular, aligned-spin version of the \texttt{TEOBResumS-Dal\'i} waveform model, thereby avoiding possible waveform systematics. In this setup, the in-plane spin components, eccentricity, and true anomaly are all fixed to zero. The injected signals and the recovery analysis employ identical mode content. The corresponding results are shown in Fig.~\ref{fig:BF_QCML_QCUL_TEOB}.
	First, we observe that the degeneracy between eccentricity-induced and microlensing-induced waveform features, as quantified by the Bayes factor $\log_{10}\mathcal{B}^\mathrm{QCML}_\mathrm{QCUL}$, increases almost monotonically with eccentricity across all binary masses considered. At the highest eccentricity value investigated, $e_{Mf\approx0.003}=0.4$, $\mathcal{H}_{\rm QCML}$ is very strongly favored over $\mathcal{H}_{\rm QCUL}$, with $\log_{10}\mathcal{B}^\mathrm{QCML}_\mathrm{QCUL}>1$ for all binary masses and mass ratios. Second, the degeneracy is stronger for lighter binaries and weakens with increasing total mass (maximum for the purple curves and minimum for the orange ones).
	This suggests that degeneracy increases with increasing number of GW cycles, which is expected.
	Third, we do not find any significant difference in $\log_{10}\mathcal{B}^\mathrm{QCML}_\mathrm{QCUL}$ between the two mass-ratio cases considered. 
	
	The degeneracy between the observational signatures of eccentricity and microlensing can be further understood by examining how well $\mathcal{H}_{\rm QCML}$ performs relative to $\mathcal{H}_{\rm QCUL}$ in recovering the injected parameters. In Fig.~\ref{fig:bias_TEOB}, we show the recovered intrinsic parameters: total mass $(M_\mathrm{tot})$, mass ratio $(q)$, and the aligned spin components $(\chi_{\rm 1z}, \chi_{\rm 2z})$. We show this comparison only for the $\{M_\mathrm{tot} = 30~{\rm M}_\odot, q = 1\}$ case, which corresponds to one of the configurations that led to the largest biases in microlensing searches, as seen in Fig.~\ref{fig:BF_QCML_QCUL_TEOB}. For all parameters, we find that the biases in the recovered values remain broadly consistent between $\mathcal{H}_{\rm QCML}$ and $\mathcal{H}_{\rm QCUL}$ up to $e \sim 0.2$, but start diverging at higher eccentricities. Overall, the biases are comparable or smaller for $\mathcal{H}_{\rm QCML}$ than for $\mathcal{H}_{\rm QCUL}$, indicating that microlensing can mimic eccentric features well enough to yield more accurate intrinsic parameter recovery, further reinforcing the degeneracy between eccentric and microlensed GW signals.
	
	Motivated by ongoing searches for microlensing signatures using Bayesian model comparison, we also analyze the injections with the quasicircular unlensed waveform model \texttt{IMRPhenomXPHM-SpinTaylor}~\cite{Khan:2018fmp, Garcia-Quiros:2020qpx, Pratten:2020ceb, Colleoni:2024knd}, a fully precessing model that employs all 15 parameters to describe a typical QCUL BBH waveform, including higher-order modes. The injected signals and recovery analyses use the same mode content. The results are shown in Fig.~\ref{fig:BF_QCML_QCUL_XPHM}. Here, we again observe a similar trend in $\log_{10}\mathcal{B}^\mathrm{QCML}_\mathrm{QCUL}$ as a function of eccentricity and total binary mass, an almost monotonic increase with eccentricity and stronger degeneracy for lower masses. However, for the $M_\mathrm{tot}=30~{\rm M}_\odot$ case, $\log_{10}\mathcal{B}^\mathrm{QCML}_\mathrm{QCUL}$ rises to $\gtrsim 8$ at $e=0.3$ and then starts to oscillate, deviating from the monotonic behavior.
	This deviation can be attributed to primarily three factors.
	First, since we consider signals starting from the time when the apastron frequency of the dominant $(\ell, |m|)=(2, 2)$ mode crosses the starting frequency $f_{\rm start}$, the duration of our injected signals decrease with increasing eccentricity values. This trend is opposite to what would occur in a more realistic setup where signals should ideally start from the time when the periastron frequency crosses $f_{\rm start}$. Consequently, in our configuration, the number of GW cycles decreases with increasing eccentricity, and we therefore do not expect $\log_{10}\mathcal{B}^\mathrm{QCML}_\mathrm{QCUL}$ to increase monotonically with $e$.
	Second, the degeneracy between eccentricity and the intrinsic parameters, particularly the precession, also becomes important, as illustrated in Fig.~\ref{fig:bias_XPHM}. There, we show the inferred intrinsic parameters (from left to right): the total mass $(M_\mathrm{tot})$, mass ratio $(q)$, effective precession spin parameter $(\chi_{\rm p})$, and effective aligned-spin parameter $(\chi_{\rm eff})$. As evident from the figure, there exists a significant degeneracy between eccentricity and precession, as well as between eccentricity and the total mass, consistent with previous studies~\cite{CalderonBustillo:2020xms, Romero-Shaw:2020thy, Lenon:2020oza, Favata:2021vhw, Romero-Shaw:2022fbf, Xu:2022zza, Hegde:2023yoz, Patterson:2024vbo}. 
	The oscillations in the inferred parameters could be one of the reasons for the oscillations seen in $\log_{10}\mathcal{B}^\mathrm{QCML}_\mathrm{QCUL}$.
	Additionally, unlike in Fig.~\ref{fig:bias_TEOB}, the biases in the inferred parameters are not always smaller for $\mathcal{H}_{\rm QCML}$ than for $\mathcal{H}_{\rm QCUL}$.
	Due to this degeneracy between eccentricity and the precession, the Bayes factors in case of \texttt{IMRPhenomXPHM-SpinTaylor} are always lower than that of \texttt{TEOBResumS-Dal\'i} for cases with $\log_{10}\mathcal{B}^\mathrm{QCML}_\mathrm{QCUL}>1$.
	Third, even if we assume that the degeneracy between microlensing and eccentric features increases monotonically with eccentricity, we should still not expect $\log_{10}\mathcal{B}^\mathrm{QCML}_\mathrm{QCUL}$ to increase monotonically with $e$. This is because the prior on $y$ favors larger values, where microlensing effects are weak (see Eq.~\ref{eq:lensing_priors_final}). Since the evidence depends sensitively on the prior volume, the stronger microlensing features at low $y$ are down-weighted by the prior. As a result, the combined effect could lead to a non-monotonic behavior of $\log_{10}\mathcal{B}^\mathrm{QCML}_\mathrm{QCUL}$ as a function of eccentricity.
	
	\begin{figure}
		\centering
		\includegraphics[width=1\linewidth]{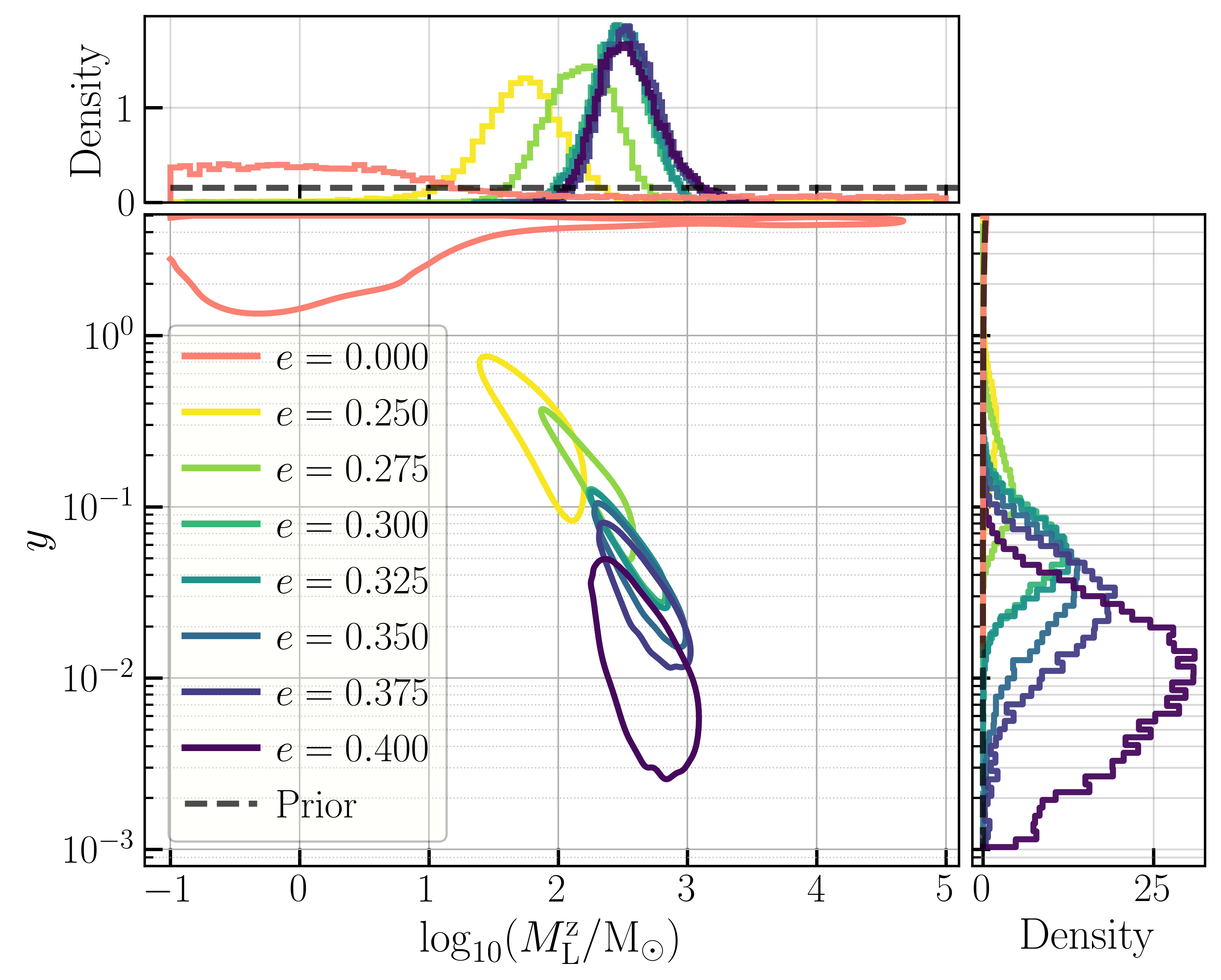}
		\caption{Inferred microlens parameters, $\log_{10}M^{\rm z}_{\rm L}$ and $y$, for non-spinning eccentric injections with $\{M_\mathrm{tot} = 30~{\rm M}_\odot, q = 1\}$, analyzed using \texttt{IMRPhenomXPHM-SpinTaylor}. We show results only for cases with $\log_{10}\mathcal{B}^\mathrm{QCML}_\mathrm{QCUL}>1$, with the $e=0$ case included for reference. The bottom-left panel displays the inferred $1\sigma$ credible regions for the lens parameters, while the adjacent panels show the corresponding one-dimensional marginalized posteriors.}
		\label{fig:MicL_Params_TEOB_Injs_1_sigma}
	\end{figure}
	
	For $\{M_\mathrm{tot} = 30~{\rm M}_\odot, q = 1\}$ injections analyzed with \texttt{IMRPhenomXPHM-SpinTaylor}, the inferred $1\sigma$ credible regions for the lens parameters, $M_\mathrm{L}^z$ and $y$, are shown in Fig.~\ref{fig:MicL_Params_TEOB_Injs_1_sigma} for all cases with $\log_{10}\mathcal{B}^\mathrm{QCML}_\mathrm{QCUL}>1$, with the $e=0$ case included for reference. As seen in the figure, for all cases where the evidence favors $\mathcal{H}_{\rm QCML}$, the posteriors for the lens parameters are well localized. The inferred $M^{\rm z}_{\rm L}$ values typically lie in the range $\sim 10$–$10^3~$M$_\odot$. Furthermore, the inferred $y$ decreases with increasing eccentricity. All these contours differ significantly from those obtained for a truly quasicircular injection (salmon-colored contour).
	
	In this section, we have demonstrated that neglecting eccentricity can introduce significant biases when searching for microlensing signatures in GW signals from BBH systems. 
	Our results also suggest that the interplay among precession, eccentricity, and microlensing may impact the analysis in a complex manner.
	Next, we investigate whether this degeneracy can be mitigated by employing waveform models consistent with the injections, that is, by doing the inference using eccentric waveform models.
	
	\begin{figure}[t]
		\centering
		\includegraphics[width=\linewidth]{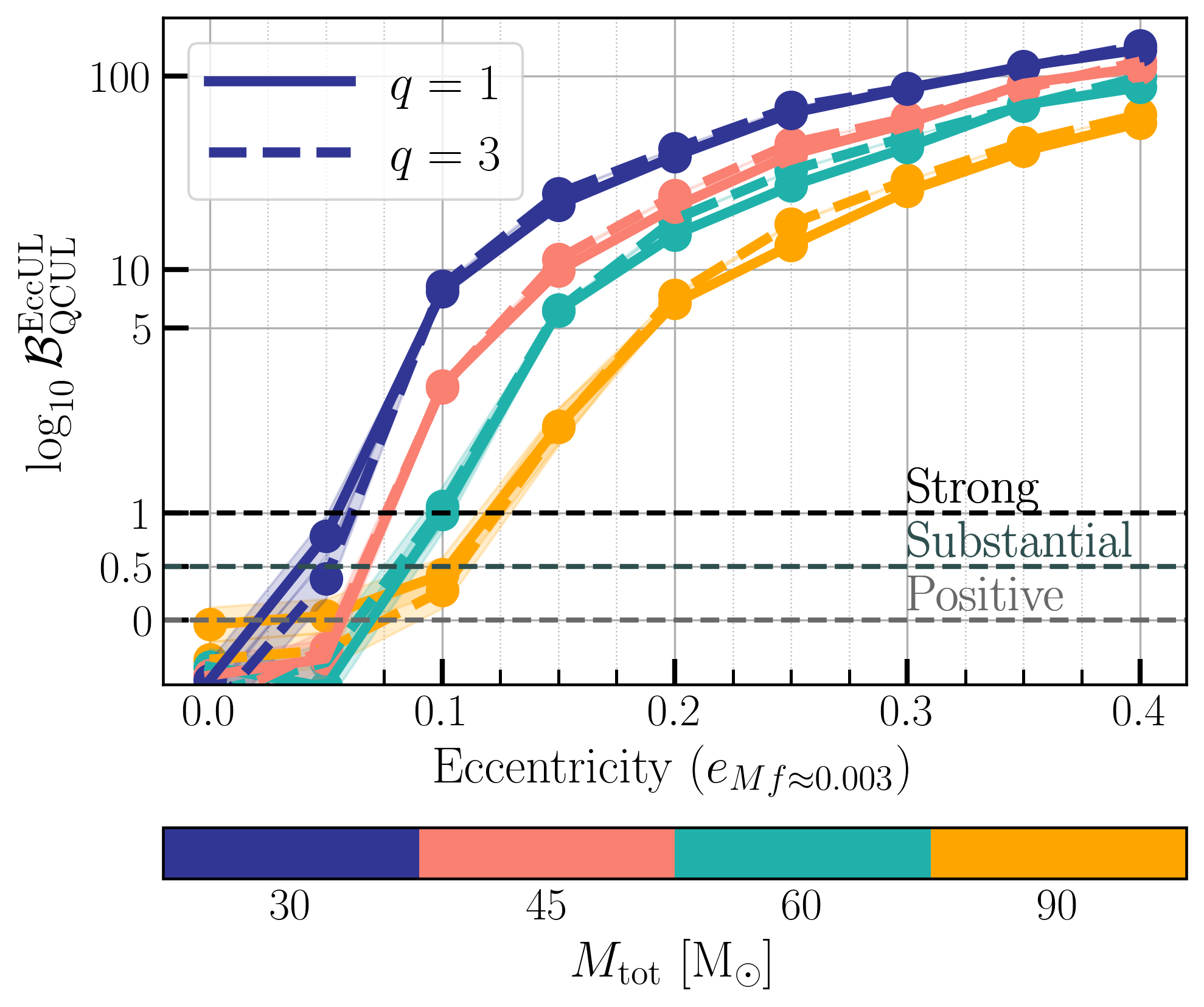}
		\caption{
			Variation in the Bayes factors between the EccUL and QCUL hypotheses, $\log_{10}\mathcal{B}^\mathrm{EccUL}_\mathrm{QCUL}$, shown as a function of the eccentricity $(e)$ (x-axis), total binary mass $(M_\mathrm{tot})$ (color scale), and mass ratio $(q)$ (solid vs.~dashed lines). Eccentricity is defined at a dimensionless frequency of $\sim 0.003$ at apastron. 
			The uncertainty in the estimated Bayes factors are represented by the translucent bands around each curve.
		}
		\label{fig:BF_EccUL_QCUL_TEOB}
	\end{figure}

	\subsubsection{\label{subsubsec:Ecc_recs}Breaking the Degeneracy: Eccentric Inference for Eccentric Injections}
	In this section, we investigate whether the degeneracy between eccentricity-induced and microlensing-induced features can be broken when analyzing eccentric injections with eccentric templates, i.e., templates that include all relevant physical effects present in the injections.
	
	We analyze the injections using the \texttt{TEOBResumS-Dal\'i} waveform model under the EccUL hypothesis, which is the same model used for the injections. The eccentricity is treated as a free parameter with a uniform prior in $[0, 0.45]$, defined at the apastron frequency, which is set equal to the waveform starting frequency of $f_{\rm start} = 10$ Hz. The true anomaly and the in-plane spin components are fixed to zero, while the aligned spin components vary in $[-0.9, 0.9]$ with a prior uniform in magnitude and isotropic in orientation (uniform in the cosine of the tilt angle). All remaining priors are chosen to be agnostic and sufficiently broad around the true injected values.
	We employ the same mode content used in the injections and set the likelihood low-frequency cutoff to $f_{\rm low}=20$ Hz. The sampling frequency is $2048$ Hz. All other settings are kept consistent with those used for the EccUL analysis in Sec.~\ref{subsec:NR_injs}.
	
	The results are shown in Fig.~\ref{fig:BF_EccUL_QCUL_TEOB}, where we display the Bayes factors between the EccUL and QCUL hypotheses, $\log_{10}\mathcal{B}^\mathrm{EccUL}_\mathrm{QCUL}$. We first observe that $\log_{10}\mathcal{B}^\mathrm{EccUL}_\mathrm{QCUL}$ increases monotonically with eccentricity across all $\{M_\mathrm{tot}, q\}$ configurations considered. Comparing these values with the corresponding $\log_{10}\mathcal{B}^\mathrm{QCML}_\mathrm{QCUL}$ in Fig.~\ref{fig:BF_QCML_QCUL_TEOB}, we find that the former becomes an order of magnitude larger than the latter for $e \gtrsim 0.05$. This demonstrates that the degeneracy between eccentricity and microlensing can indeed be broken when the analysis employs waveform models that correctly capture all relevant physics of the signal, namely eccentricity, preventing erroneous claims of microlensing.

	\section{\label{sec:conclusion}Conclusion}
	In this work, we investigated the extent to which orbital eccentricity in BBH GW signals can mimic the observational signatures of microlensing produced by an isolated point-mass lens, potentially leading to false claims of microlensing.
	Using a combination of highly reliable NR simulations, \texttt{TEOBResumS-Dal\'i}eccentric injections, and Bayesian model comparison analyses, supported by a complementary mismatch study across a simulated eccentric BBH population, we assessed when this degeneracy becomes significant and whether it can be resolved by including eccentric waveform models in the recovery.
	
	Our main conclusions are as follows:
	\begin{enumerate}
		\item Bayesian model comparison analyses of both NR and \texttt{TEOBResumS-Dal\'i} EccUL injections demonstrate that ignoring eccentricity can bias microlensing searches in certain regions of the parameter space. Specifically, this degeneracy strengthens with (i) increasing eccentricity, (ii) decreasing total mass, and (iii) increasing SNR. Consequently, $\mathcal{H}_{\rm QCML}$ can be strongly favored over $\mathcal{H}_{\rm QCUL}$ under these conditions, even when the true signal is not microlensed.
		
		\item For NR injections with a gauge-independent eccentricity measured at $20~\mathrm{Hz}$ to be $e^{\mathrm{gw}}_{20~\mathrm{Hz}}\sim 0.3$, total mass $M_\mathrm{tot}\approx 80~$M$_\odot$, and mass ratios $q\in\{1,2,3\}$, we find that the $\mathcal{H}_{\rm QCML}$ is very strongly preferred over the $\mathcal{H}_{\rm QCUL}$, yielding $\log_{10}\mathcal{B}^\mathrm{QCML}_\mathrm{QCUL}\gtrsim 10$,  when the true injected optimal network SNR is around $100$. In contrast, no such bias appears for lower-SNR injections ($\sim 15$), for which $\log_{10}\mathcal{B}^\mathrm{QCML}_\mathrm{QCUL}<0$ across all eccentric injections.
		
		\item For a network optimal SNR of $30$, binaries with $M_\mathrm{tot}\lesssim 100~$M$_\odot$ and high eccentricities ($\gtrsim 0.4$, defined at apastron at $Mf\approx0.003$), almost always favor the $\mathcal{H}_{\rm QCML}$, having $\log_{10}\mathcal{B}^\mathrm{QCML}_\mathrm{QCUL}\gtrsim 1$. 
		
		\item In cases where the $\mathcal{H}_{\rm QCML}$ is strongly favored over the $\mathcal{H}_{\rm QCUL}$, the inferred microlens parameters, $\log_{10}M^{\rm z}_{\rm L}$ and $y$, exhibit well-localized posteriors and large Jensen-Shannon divergence values between the priors and the inferred 1D marginalized posteriors for the lens parameters. 
		
		\item  Since $\mathcal{H}_{\rm QCUL}$ is a subset of $\mathcal{H}_{\rm QCML}$ (because microlensing effects are added on top of the unlensed waveform, in the weak microlensing regime both hypotheses produce similar, or \textit{faithful}, signals), we expect $\mathcal{H}_{\rm QCML}$ to perform at least as well as $\mathcal{H}_{\rm QCUL}$ when the true signal is EccUL (or, more generally, when any physics is missing from the model). However, introducing additional parameters in $\mathcal{H}_{\rm QCML}$ also incurs a penalty owing to \textit{Occam's razor}, which suppresses false microlensing support except in the regimes identified above. 
		
		\item For both NR and \texttt{TEOBResumS-Dal\'i} EccUL injections, the degeneracy is always broken when the recovery uses eccentric templates containing the correct physics. In such cases, $\log_{10}\mathcal{B}^\mathrm{EccUL}_\mathrm{QCUL}\gg \log_{10}\mathcal{B}^\mathrm{QCML}_\mathrm{QCUL}$, typically by an order of magnitude. Thus, events which are prone to showing strong, but spurious, support for microlensing are expected to show even greater support for eccentricity when recovered with an appropriate eccentric waveform model. Thus, the degeneracy is not intrinsic to the signal: it reflects an incomplete model space rather than any physical ambiguity. 
		
	\end{enumerate}
	
	
	The findings suggest that although $\mathcal{H}_{\rm QCML}$ can be strongly favored over the $\mathcal{H}_{\rm QCUL}$, it occurs only when either the signal is strongly eccentric or has a high enough SNR. The broader implications of this degeneracy are twofold. First, any event showing strong evidence for microlensing under a quasicircular model must be subjected to a parallel analysis using eccentric waveform families. Only if the microlensing hypothesis remains preferred once eccentricity is allowed can a physical lens interpretation be considered robust. Second, since the effects of non-zero eccentricity become increasingly important when probing earlier stages of binary evolution or with improved detector sensitivities, our results carry important implications for future observatories such as third-generation detectors (CE, ET) and space-based missions like DECIGO and LISA. At lower frequencies, eccentric modulations persist for more cycles and can more readily imitate the diffraction patterns of microlensing, while higher SNRs will make any spurious preference statistically more significant.
	
	However, we emphasize that many astrophysical formation channels that produce substantial eccentricities, including dynamical interactions and hierarchical triples, often generate significant in-plane spins and hence strong precession.
	The current lack of \textit{faithful} precessing-eccentric waveform models may, therefore, pose challenges to fully breaking this degeneracy in realistic analyses. In such scenarios, models incorporating both microlensing and precession may outperform aligned-spin eccentric models. We will explore this interesting line of investigation in our future work.
	
	Our findings highlight a clear prescription for microlensing searches: strong Bayesian support for microlensing under quasicircular models does not guarantee a physical lensing event, and must always be tested for other \emph{atypical} physical effects, such as eccentricity, to ensure unbiased, astrophysically reliable interpretation in the era of precision GW astronomy. Finally, while we focused here primarily on biases in microlensing searches introduced by ignoring non-zero eccentricity, the converse possibility is equally important: biases in eccentricity searches induced by microlensing or other atypical physical effects. Such degeneracies may be resolved by analyzing the evolution of the inferred eccentricity as a function of frequency and assessing its consistency with post-Newtonian predictions~\cite{Bhat:2025lri}.
	Investigating such reverse degeneracies will further clarify the robustness of eccentricity inference in future GW observations.
	
	\begin{figure*}[t]
		\centering
		\begin{subfigure}[t]{0.49\textwidth}
			\centering
			\includegraphics[width=\linewidth]{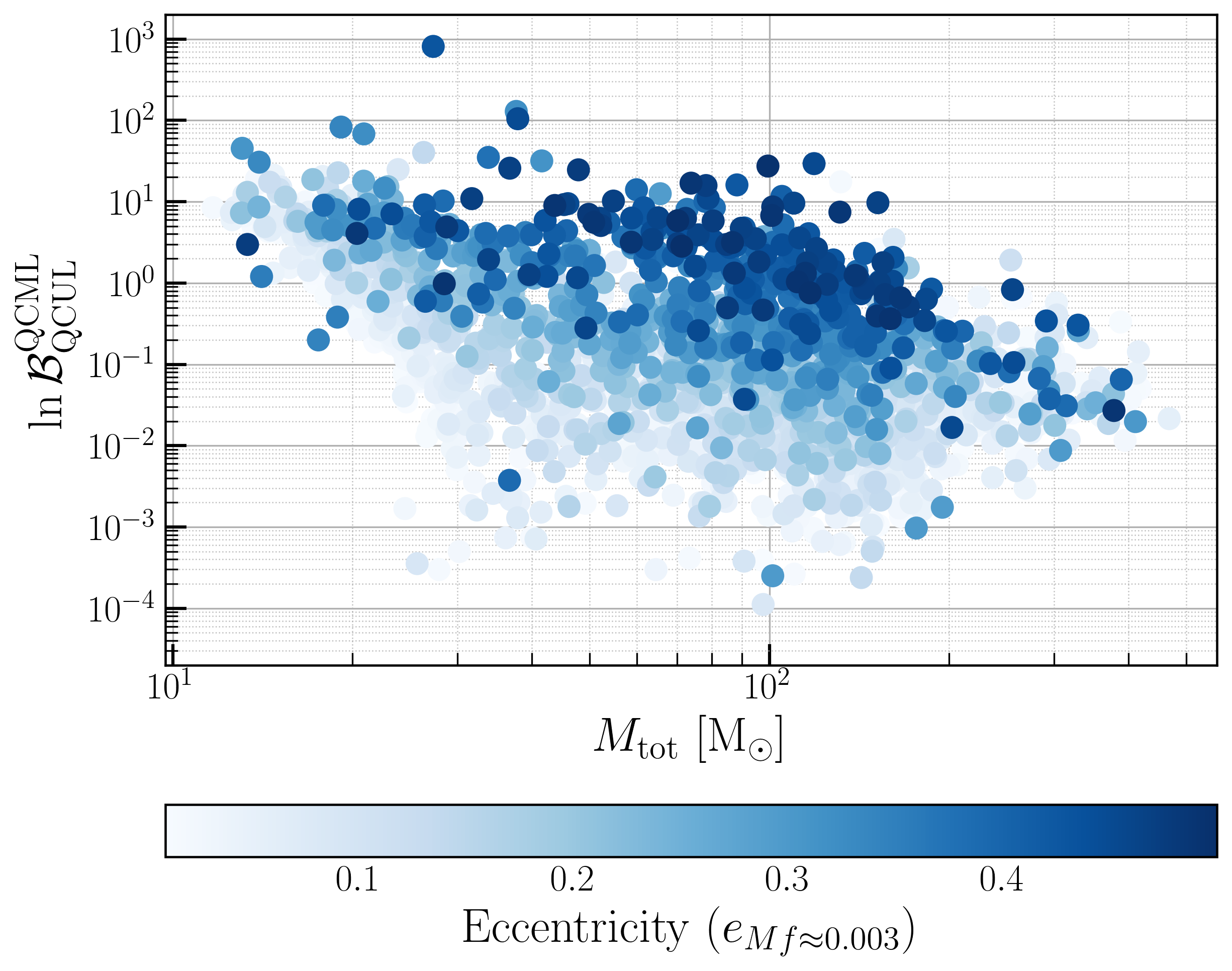}
			\label{fig:corr_M_BF}
		\end{subfigure}
		\hfill
		\begin{subfigure}[t]{0.49\textwidth}
			\centering
			\includegraphics[width=\linewidth]{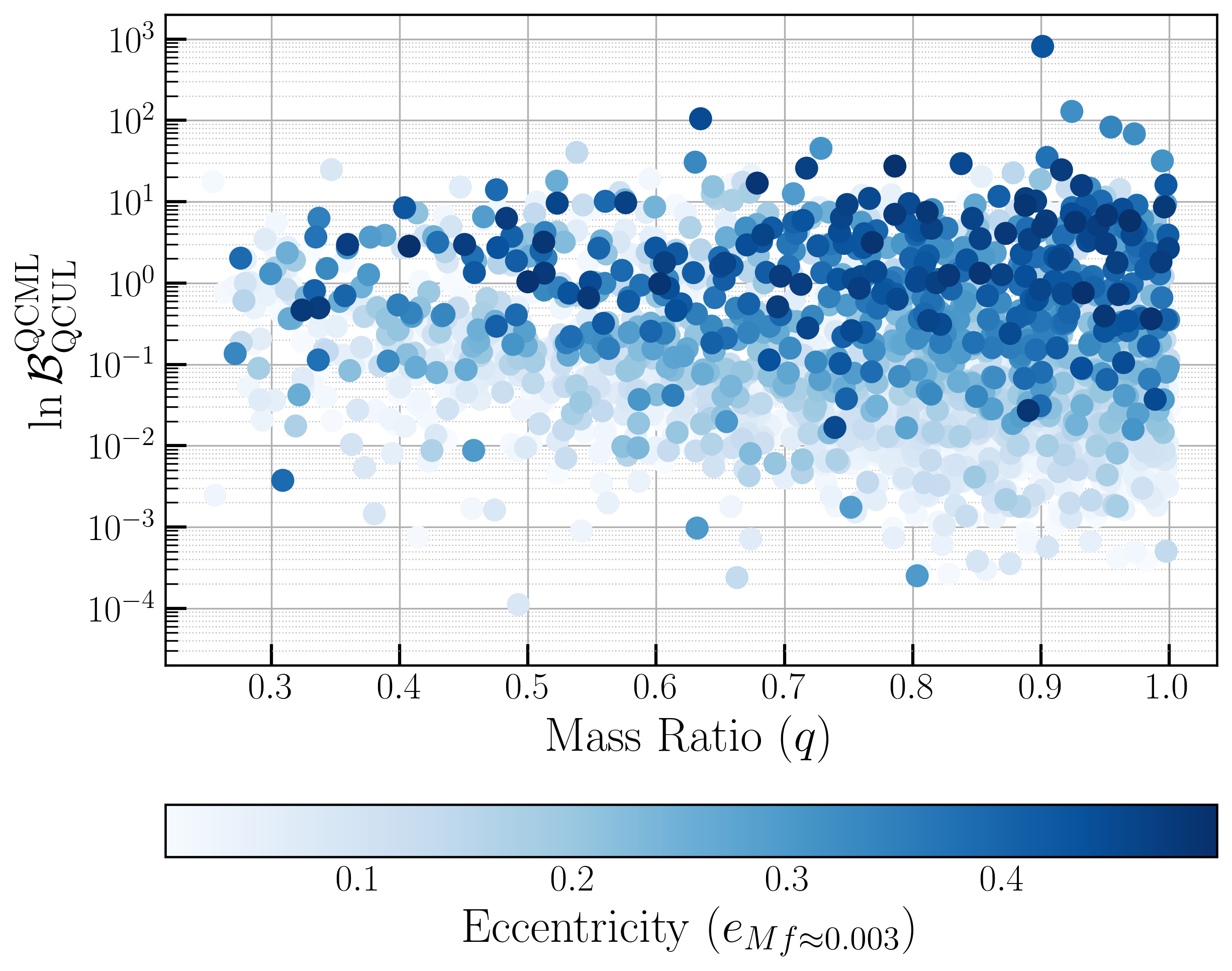}
			\label{fig:corr_q_BF}
		\end{subfigure}
		
		\caption{$\ln\mathcal{B}^\mathrm{QCML}_\mathrm{QCUL}$ estimated from fitting factor differences using Eq.~\ref{eq:BF_estimate_using_FF} for the simulated eccentric BBH population, plotted against total mass $M_\mathrm{total}$ (left) and mass ratio $q$ (right). The color bar denotes eccentricity ($e_{M f \approx 0.003}$). $\ln\mathcal{B}^\mathrm{QCML}_\mathrm{QCUL}$ increases with eccentricity, shows a slight anti-correlation with total mass, and exhibits no significant trend with mass ratio.}
		\label{fig:Ecc_pop_FF_study}
	\end{figure*}
	
	\begin{acknowledgments}
		We sincerely thank Nathan K. Johnson-McDaniel for many useful discussions during the early phase of this project. We also thank Akash Maurya for many useful comments and discussions, and Aditya Vijaykumar for carefully reading the manuscript.
		We would also like to thank P. Ajith, Sukanta Bose, Anuradha Gupta, Shasvath Kapadia, Prayush Kumar, Paul Lasky, Anupreeta More, Purnima Narayan, Anirudh S. Nemmani, Archana Pai, Arif Shaikh, and Aditya Sharma for their useful comments and questions.
		The authors are grateful for computational resources provided by the LIGO Laboratory and supported by National Science Foundation Grants PHY-0757058 and PHY-0823459. We also acknowledge the use of the IUCAA LDG cluster Sarathi for computational work.
		
		The work utilizes the following software packages:
		\texttt{Cython}~\cite{behnel2011cython},
		\texttt{NumPy}~\cite{Harris:2020xlr}, 
		\texttt{SciPy}~\cite{Virtanen:2019joe}, 
		\texttt{PyCBC}~\cite{pycbc_github},
		\texttt{LALSuite}~\cite{2020ascl.soft12021L},
		\texttt{dynesty}~\cite{2020MNRAS.493.3132S}, 
		\texttt{Bilby}~\cite{Ashton:2018jfp,Smith:2019ucc}, 
		\texttt{PESummary}~\cite{Hoy:2020vys},
		\texttt{gw\_eccentricity}~\cite{Shaikh:2023ypz, Shaikh:2025tae},
		\texttt{GWPopulation}~\cite{2019PhRvD.100d3030T}, 
		\texttt{Matplotlib}~\cite{Hunter:2007}, and
		\texttt{Jupyter notebook}~\cite{Kluyver2016jupyter}.
		
	\end{acknowledgments} 
	
	\appendix
	\section{\label{subsec:Ecc_pop}Microlensing biases for an Eccentric BBH Population using Mismatch Analysis}
	To understand the broader impact of eccentricity on microlensing searches, we study a population of eccentric BBH signals. We generate mock GW data of around $\sim 2\times10^4$ non-spinning eccentric BBH signals using \texttt{TEOBResumS-Dal\'i}~\cite{Damour:2014sva, Nagar:2015xqa, Nagar:2018zoe, Nagar:2019wds, Nagar:2020pcj, Riemenschneider:2021ppj, Nagar:2023zxh} waveform model. The mass and spin priors are derived from the inferred population model based on GWTC-3 data~\cite{KAGRA:2021duu}: the source-frame component masses are drawn from the \textsc{PowerLaw+Peak} distribution, and spins are assumed to be independent and identically distributed, with magnitudes following a Beta distribution and tilt angles modeled as an isotropic plus truncated half-Gaussian mixture~\cite{Talbot:2019okv}. The merger rate density follow the Madau$-$Dickinson profile~\cite{Madau:2014bja, Fishbach:2018edt}, with cosmological parameters taken from with~\cite{Planck:2015fie}. 
	For simplicity, we only consider one detector (Hanford LIGO detector) for this study, and consider only those signals that have an SNR above $8$. The noise PSD correspond to the target O4 LIGO sensitivity.
	We define eccentricities at a fixed point in the binary evolution by imposing a dimensionless frequency of $\sim 0.003$ at apastron, equivalent to $10$ Hz for a $60~$M$_\odot$ BBH. The eccentricity prior is log-uniform over $e\in (0.01,0.5)$.
	
	Our goal is to compare the performance of $\mathcal{H}_{\rm QCUL}$ and $\mathcal{H}_{\rm QCML}$ when analyzing eccentric signals. Since computing evidences with a nested sampler is computationally expensive, we utilize Laplace's approximation as used in Eq.~26 of \cite{Cornish:2011ys} (neglecting the Occam factor) to approximate the Bayes factor of a hypothesis $\mathcal{H}$ against the noise-only hypothesis:
	\begin{equation}
		\ln\mathcal{B}^{\rm \mathcal{H}}_{\rm noise} \approx \frac{\rho_{\mathcal{H}}^2}{2}
	\end{equation}
	
	Thus, $\ln\mathcal{B}^\mathrm{QCML}_\mathrm{QCUL}$ can be estimated as:
	\begin{equation}
		\begin{aligned}
			\ln\mathcal{B}^\mathrm{QCML}_\mathrm{QCUL} &= \ln\mathcal{B}^\mathrm{QCML}_\mathrm{noise} - \ln\mathcal{B}^\mathrm{QCUL}_\mathrm{noise}  \\
			&=  \frac{1}{2} (\rho_{\mathcal{H_{\rm QCML}}}^2 -\rho_{\mathcal{H_{\rm QCUL}}}^2) \\
			&= \frac{1}{2} (FF_{\rm QCML}^2 - FF_{\rm QCUL}^2)\rho_{\rm EccUL}^2 \\
		\end{aligned}
		\label{eq:BF_estimate_using_FF}
	\end{equation}
	
	where $FF_\mathrm{QCUL}$ ($FF_\mathrm{QCML}$) are the fitting factor (FF) values when QCUL (QCML) model is used to analyze the EccUL target waveforms, and $\rho_\mathrm{EccUL}$ depicts the true optimal SNR of the eccentric signal. In writing this, we used the relation $\rho_\mathrm{EccUL} = \rho_\mathrm{QCUL}/FF_\mathrm{QCUL} = \rho_\mathrm{QCML}/FF_\mathrm{QCML}$~\cite{Ajith:2012mn}.
	We use \texttt{TEOBResumS-Dal\'i} for computing \textit{match} values, with the lower and upper frequency cutoffs set to $20~$Hz and $1024~$Hz, respectively. To compute the FF values, we use the Nelder-Mead algorithm~\cite{Nelder:1965zz}, as implemented in the \texttt{optimization} module of the \texttt{Scipy} library \cite{Virtanen:2019joe}, to maximize the \textit{match} between the waveforms (for more details, see Sec.~2.2 of \cite{Mishra:2023ddt}). For $\mathcal{H}_{\rm QCUL}$, we employ a 4D aligned-spin waveform (WF) model to analyze the eccentric WFs, parameterized by: chirp mass $(\mathcal{M}_c)$, mass ratio $(q)$, and aligned spin components of the two component masses $(\chi_{1z},~\chi_{2z})$. For $\mathcal{H}_{\rm QCML}$, we use a 6D WF model with two additional parameters than that in $\mathcal{H}_{\rm QCUL}$, namely, $\log_{10}M^{\rm z}_{\rm L}$ and $y$.
	All remaining parameters are fixed to their true values. Since we maximize over continuous parameters, our FF computation does not include additional loss from a discrete \textit{template bank}; we isolate only the loss due to \textit{missing physics} in the waveform models used for the inference. 
	
	In Fig.~\ref{fig:Ecc_pop_FF_study}, we show $\ln\mathcal{B}^\mathrm{QCML}_\mathrm{QCUL}$ for our simulated eccentric BBH population against the total binary mass (left panel) and mass ratio (right panel), with color bar representing eccentricity values. 
	We observe: (i) For any value on the x-axis, the $\ln\mathcal{B}^\mathrm{QCML}_\mathrm{QCUL}$ increases with increasing eccentricity values (i.e., the points transition from lighter to darker shades as the y-values rise); (ii) For a fixed eccentricity value, the left panel of Fig.\ref{fig:Ecc_pop_FF_study} shows a mild anti-correlation between $\ln\mathcal{B}^\mathrm{QCML}_\mathrm{QCUL}$ and the total binary mass $M_\mathrm{total}$; and (iii) the right panel of Fig.~\ref{fig:Ecc_pop_FF_study} shows no significant correlation between $\ln\mathcal{B}^\mathrm{QCML}_\mathrm{QCUL}$ and $q$.
	
	Overall, the fitting factor analysis presented in this section is consistent with the full Bayesian model-comparison results of Sec.~\ref{sec:res}.
	
	\bibliography{main} 

@article{LIGOScientific:2014pky,
    author = "Aasi, J. and others",
    collaboration = "LIGO Scientific",
    title = "{Advanced LIGO}",
    eprint = "1411.4547",
    archivePrefix = "arXiv",
    primaryClass = "gr-qc",
    doi = "10.1088/0264-9381/32/7/074001",
    journal = "Class. Quant. Grav.",
    volume = "32",
    pages = "074001",
    year = "2015"
}

@article{LIGOScientific:2021izm,
    author = "Abbott, R. and others",
    collaboration = "LIGO Scientific, VIRGO",
    title = "{Search for Lensing Signatures in the Gravitational-Wave Observations from the First Half of LIGO\textendash{}Virgo\textquoteright{}s Third Observing Run}",
    eprint = "2105.06384",
    archivePrefix = "arXiv",
    primaryClass = "gr-qc",
    reportNumber = "LIGO-P2000400",
    doi = "10.3847/1538-4357/ac23db",
    journal = "Astrophys. J.",
    volume = "923",
    number = "1",
    pages = "14",
    year = "2021"
}

@article{LIGOScientific:2018mvr,
    author = "Abbott, B. P. and others",
    collaboration = "LIGO Scientific, Virgo",
    title = "{GWTC-1: A Gravitational-Wave Transient Catalog of Compact Binary Mergers Observed by LIGO and Virgo during the First and Second Observing Runs}",
    eprint = "1811.12907",
    archivePrefix = "arXiv",
    primaryClass = "astro-ph.HE",
    reportNumber = "LIGO-P1800307",
    doi = "10.1103/PhysRevX.9.031040",
    journal = "Phys. Rev. X",
    volume = "9",
    number = "3",
    pages = "031040",
    year = "2019"
}

@article{LIGOScientific:2020ibl,
    author = "Abbott, R. and others",
    collaboration = "LIGO Scientific, Virgo",
    title = "{GWTC-2: Compact Binary Coalescences Observed by LIGO and Virgo During the First Half of the Third Observing Run}",
    eprint = "2010.14527",
    archivePrefix = "arXiv",
    primaryClass = "gr-qc",
    reportNumber = "P2000061",
    doi = "10.1103/PhysRevX.11.021053",
    journal = "Phys. Rev. X",
    volume = "11",
    pages = "021053",
    year = "2021"
}

@article{LIGOScientific:2021usb,
    author = "Abbott, R. and others",
    collaboration = "LIGO Scientific, VIRGO",
    title = "{GWTC-2.1: Deep extended catalog of compact binary coalescences observed by LIGO and Virgo during the first half of the third observing run}",
    eprint = "2108.01045",
    archivePrefix = "arXiv",
    primaryClass = "gr-qc",
    reportNumber = "LIGO-P2100063",
    doi = "10.1103/PhysRevD.109.022001",
    journal = "Phys. Rev. D",
    volume = "109",
    number = "2",
    pages = "022001",
    year = "2024"
}

@article{LIGOScientific:2025slb,
    author = "Abac, A. G. and others",
    collaboration = "LIGO Scientific, VIRGO, KAGRA",
    title = "{GWTC-4.0: Updating the Gravitational-Wave Transient Catalog with Observations from the First Part of the Fourth LIGO-Virgo-KAGRA Observing Run}",
    journal = "arXiv:2508.18082",
    eprint = "2508.18082",
    archivePrefix = "arXiv",
    primaryClass = "gr-qc",
    reportNumber = "LIGO-P2400386",
    month = "8",
    year = "2025"
}

@article{LIGOScientific:2016wof,
    author = "Abbott, Benjamin P and others",
    collaboration = "LIGO Scientific",
    title = "{Exploring the Sensitivity of Next Generation Gravitational Wave Detectors}",
    eprint = "1607.08697",
    archivePrefix = "arXiv",
    primaryClass = "astro-ph.IM",
    reportNumber = "LIGO-P1600143",
    doi = "10.1088/1361-6382/aa51f4",
    journal = "Class. Quant. Grav.",
    volume = "34",
    number = "4",
    pages = "044001",
    year = "2017"
}

@article{VIRGO:2014yos,
    author = "Acernese, F. and others",
    collaboration = "VIRGO",
    title = "{Advanced Virgo: a second-generation interferometric gravitational wave detector}",
    eprint = "1408.3978",
    archivePrefix = "arXiv",
    primaryClass = "gr-qc",
    doi = "10.1088/0264-9381/32/2/024001",
    journal = "Class. Quant. Grav.",
    volume = "32",
    number = "2",
    pages = "024001",
    year = "2015"
}

@article{Somiya:2011np,
    author = "Somiya, Kentaro",
    editor = "Hannam, Mark and Sutton, Patrick and Hild, Stefan and van den Broeck, Chris",
    collaboration = "KAGRA",
    title = "{Detector configuration of KAGRA: The Japanese cryogenic gravitational-wave detector}",
    eprint = "1111.7185",
    archivePrefix = "arXiv",
    primaryClass = "gr-qc",
    doi = "10.1088/0264-9381/29/12/124007",
    journal = "Class. Quant. Grav.",
    volume = "29",
    pages = "124007",
    year = "2012"
}

@article{Aso:2013eba,
    author = "Aso, Yoichi and Michimura, Yuta and Somiya, Kentaro and Ando, Masaki and Miyakawa, Osamu and Sekiguchi, Takanori and Tatsumi, Daisuke and Yamamoto, Hiroaki",
    collaboration = "KAGRA",
    title = "{Interferometer design of the KAGRA gravitational wave detector}",
    eprint = "1306.6747",
    archivePrefix = "arXiv",
    primaryClass = "gr-qc",
    doi = "10.1103/PhysRevD.88.043007",
    journal = "Phys. Rev. D",
    volume = "88",
    number = "4",
    pages = "043007",
    year = "2013"
}

@article{KAGRA:2018plz,
    author = "Akutsu, T. and others",
    collaboration = "KAGRA",
    title = "{KAGRA: 2.5 Generation Interferometric Gravitational Wave Detector}",
    eprint = "1811.08079",
    archivePrefix = "arXiv",
    primaryClass = "gr-qc",
    reportNumber = "JGW-P1809243",
    doi = "10.1038/s41550-018-0658-y",
    journal = "Nature Astron.",
    volume = "3",
    number = "1",
    pages = "35--40",
    year = "2019"
}

@article{KAGRA:2020tym,
    author = "Akutsu, T. and others",
    collaboration = "KAGRA",
    title = "{Overview of KAGRA: Detector design and construction history}",
    eprint = "2005.05574",
    archivePrefix = "arXiv",
    primaryClass = "physics.ins-det",
    doi = "10.1093/ptep/ptaa125",
    journal = "PTEP",
    volume = "2021",
    number = "5",
    pages = "05A101",
    year = "2021"
}

@article{KAGRA:2021vkt,
    author = "Abbott, R. and others",
    collaboration = "KAGRA, VIRGO, LIGO Scientific",
    title = "{GWTC-3: Compact Binary Coalescences Observed by LIGO and Virgo during the Second Part of the Third Observing Run}",
    eprint = "2111.03606",
    archivePrefix = "arXiv",
    primaryClass = "gr-qc",
    reportNumber = "LIGO-P2000318",
    doi = "10.1103/PhysRevX.13.041039",
    journal = "Phys. Rev. X",
    volume = "13",
    number = "4",
    pages = "041039",
    year = "2023"
}

@article{Punturo:2010zz,
    author = "Punturo, M. and others",
    editor = "Ricci, Fulvio",
    title = "{The Einstein Telescope: A third-generation gravitational wave observatory}",
    doi = "10.1088/0264-9381/27/19/194002",
    journal = "Class. Quant. Grav.",
    volume = "27",
    pages = "194002",
    year = "2010"
}

@article{Hild:2010id,
    author = "Hild, S. and others",
    title = "{Sensitivity Studies for Third-Generation Gravitational Wave Observatories}",
    eprint = "1012.0908",
    archivePrefix = "arXiv",
    primaryClass = "gr-qc",
    doi = "10.1088/0264-9381/28/9/094013",
    journal = "Class. Quant. Grav.",
    volume = "28",
    pages = "094013",
    year = "2011"
}

@article{Reitze:2019iox,
    author = "Reitze, David and others",
    title = "{Cosmic Explorer: The U.S. Contribution to Gravitational-Wave Astronomy beyond LIGO}",
    eprint = "1907.04833",
    archivePrefix = "arXiv",
    primaryClass = "astro-ph.IM",
    reportNumber = "LIGO-P1900316",
    journal = "Bull. Am. Astron. Soc.",
    volume = "51",
    number = "7",
    pages = "035",
    year = "2019"
}

@article{Regimbau:2016ike,
    author = "Regimbau, T. and Evans, M. and Christensen, N. and Katsavounidis, E. and Sathyaprakash, B. and Vitale, S.",
    title = "{Digging deeper: Observing primordial gravitational waves below the binary black hole produced stochastic background}",
    eprint = "1611.08943",
    archivePrefix = "arXiv",
    primaryClass = "astro-ph.CO",
    doi = "10.1103/PhysRevLett.118.151105",
    journal = "Phys. Rev. Lett.",
    volume = "118",
    number = "15",
    pages = "151105",
    year = "2017"
}

@article{LISA:2017,
    author = {Amaro-Seoane, Pau and others},
    title = "{Laser Interferometer Space Antenna}",
    journal = "arXiv:1702.00786",
    eprint = "1702.00786",
    archivePrefix = "arXiv",
    primaryClass = "astro-ph.IM",
    year = "2017"
}

@article{Kawamura:2006up,
    author = "Kawamura, S. and others",
    editor = "Mio, N.",
    title = "{The Japanese space gravitational wave antenna DECIGO}",
    doi = "10.1088/0264-9381/23/8/S17",
    journal = "Class. Quant. Grav.",
    volume = "23",
    pages = "S125--S132",
    year = "2006"
}

@article{Einstein:1936llh,
    author = "Einstein, Albert",
    title = "{Lens-Like Action of a Star by the Deviation of Light in the Gravitational Field}",
    doi = "10.1126/science.84.2188.506",
    journal = "Science",
    volume = "84",
    pages = "506--507",
    year = "1936"
}

@article{Zwicky:1937zzb,
    author = "Zwicky, F.",
    title = "{Nebulae as gravitational lenses}",
    doi = "10.1103/PhysRev.51.290",
    journal = "Phys. Rev.",
    volume = "51",
    pages = "290",
    year = "1937"
}

@ARTICLE{1988A&A...191L..19S,
       author = {{Soucail}, G. and {Mellier}, Y. and {Fort}, B. and {Mathez}, G. and {Cailloux}, M.},
        title = "{The giant arc in A 370 : spectroscopic evidence for gravitational lensing from a source at Z=0.724.}",
      journal = {Astronomy and Astrophysics},
         year = 1988,
        month = feb,
       volume = {191},
        pages = {L19-L21},
       adsurl = {https://ui.adsabs.harvard.edu/abs/1988A&A...191L..19S}
}

@article{Walsh:1979nx,
    author = "Walsh, D. and Carswell, R. F. and Weymann, R. J.",
    title = "{0957 + 561 A, B - Twin quasistellar objects or gravitational lens}",
    doi = "10.1038/279381a0",
    journal = "Nature",
    volume = "279",
    pages = "381--384",
    year = "1979"
}

@ARTICLE{1971NCimB...6..225L,
       author = "Lawrence, J. K.",
        title = "{Focusing of gravitational radiation by interior gravitational fields.}",
      journal = {Nuovo Cimento B Serie},
         year = 1971,
        month = jan,
       volume = {6B},
        pages = {225-235},
          doi = {10.1007/BF02735388},
       adsurl = {https://ui.adsabs.harvard.edu/abs/1971NCimB...6..225L}
}

@article{Ohanian:1974ys,
    author = "Ohanian, H. C.",
    title = "{On the focusing of gravitational radiation}",
    doi = "10.1007/BF01810927",
    journal = "Int. J. Theor. Phys.",
    volume = "9",
    pages = "425--437",
    year = "1974"
}

@article{Bernardeau:1999mh,
    author = "Bernardeau, F.",
    title = "{Gravitational lenses}",
    journal = "arXiv:astro-ph/9901117",
    eprint = "astro-ph/9901117",
    archivePrefix = "arXiv",
    month = "1",
    year = "1999"
}

@article{Takahashi:2016jom,
    author = "Takahashi, Ryuichi",
    title = "{Arrival time differences between gravitational waves and electromagnetic signals due to gravitational lensing}",
    eprint = "1606.00458",
    archivePrefix = "arXiv",
    primaryClass = "astro-ph.CO",
    doi = "10.3847/1538-4357/835/1/103",
    journal = "Astrophys. J.",
    volume = "835",
    number = "1",
    pages = "103",
    year = "2017"
}

@article{Smith:2017jdz,
    author = "Smith, G. P. and others",
    editor = "Gonz\'alez, Gabriela and Hynes, Robert",
    title = "{Strong-lensing of Gravitational Waves by Galaxy Clusters}",
    eprint = "1803.07851",
    archivePrefix = "arXiv",
    primaryClass = "astro-ph.CO",
    doi = "10.1017/S1743921318003757",
    journal = "IAU Symp.",
    volume = "338",
    pages = "98--102",
    year = "2017"
}

@article{Dai:2017huk,
    author = "Dai, Liang and Venumadhav, Tejaswi",
    title = "{On the waveforms of gravitationally lensed gravitational waves}",
    journal = "arXiv:1702,04724",
    eprint = "1702.04724",
    archivePrefix = "arXiv",
    primaryClass = "gr-qc",
    month = "2",
    year = "2017"
}

@article{Haris:2018vmn,
    author = "Haris, K. and Mehta, Ajit Kumar and Kumar, Sumit and Venumadhav, Tejaswi and Ajith, Parameswaran",
    title = "{Identifying strongly lensed gravitational wave signals from binary black hole mergers}",
    journal = "arXiv:1807.07062",
    eprint = "1807.07062",
    archivePrefix = "arXiv",
    primaryClass = "gr-qc",
    reportNumber = "LIGO- P1800155",
    month = "7",
    year = "2018"
}

@article{Li:2018prc,
    author = "Li, Shun-Sheng and Mao, Shude and Zhao, Yuetong and Lu, Youjun",
    title = "{Gravitational lensing of gravitational waves: A statistical perspective}",
    eprint = "1802.05089",
    archivePrefix = "arXiv",
    primaryClass = "astro-ph.CO",
    doi = "10.1093/mnras/sty411",
    journal = "Mon. Not. Roy. Astron. Soc.",
    volume = "476",
    number = "2",
    pages = "2220--2229",
    year = "2018"
}

@article{Oguri:2018muv,
    author = "Oguri, Masamune",
    title = "{Effect of gravitational lensing on the distribution of gravitational waves from distant binary black hole mergers}",
    eprint = "1807.02584",
    archivePrefix = "arXiv",
    primaryClass = "astro-ph.CO",
    doi = "10.1093/mnras/sty2145",
    journal = "Mon. Not. Roy. Astron. Soc.",
    volume = "480",
    number = "3",
    pages = "3842--3855",
    year = "2018"
}

@article{Broadhurst:2018saj,
    author = "Broadhurst, Tom and Diego, Jose M. and Smoot, George",
    title = "{Reinterpreting Low Frequency LIGO/Virgo Events as Magnified Stellar-Mass Black Holes at Cosmological Distances}",
    journal = "arXiv:1802.05273",
    eprint = "1802.05273",
    archivePrefix = "arXiv",
    primaryClass = "astro-ph.CO",
    month = "2",
    year = "2018"
}

@article{Broadhurst:2019ijv,
    author = "Broadhurst, Tom and Diego, Jose M. and Smoot, George F.",
    title = "{Twin LIGO/Virgo Detections of a Viable Gravitationally-Lensed Black Hole Merger}",
    journal = "arXiv:1901.03190",
    eprint = "1901.03190",
    archivePrefix = "arXiv",
    primaryClass = "astro-ph.CO",
    month = "1",
    year = "2019"
}

@article{Broadhurst:2020cvm,
    author = "Broadhurst, Tom and Diego, Jose M. and Smoot, George F.",
    title = "{Interpreting LIGO/Virgo ''Mass-Gap'' events as lensed Neutron Star-Black Hole binaries}",
    journal = "arXiv:2006.13219",
    eprint = "2006.13219",
    archivePrefix = "arXiv",
    primaryClass = "astro-ph.CO",
    month = "6",
    year = "2020"
}

@article{Dai:2020tpj,
    author = "Dai, Liang and Zackay, Barak and Venumadhav, Tejaswi and Roulet, Javier and Zaldarriaga, Matias",
    title = "{Search for Lensed Gravitational Waves Including Morse Phase Information: An Intriguing Candidate in O2}",
    journal = "arXiv:2007.12709",
    eprint = "2007.12709",
    archivePrefix = "arXiv",
    primaryClass = "astro-ph.HE",
    month = "7",
    year = "2020"
}

@article{Ezquiaga:2020gdt,
    author = "Ezquiaga, Jose Mar\'\i{}a and Holz, Daniel E. and Hu, Wayne and Lagos, Macarena and Wald, Robert M.",
    title = "{Phase effects from strong gravitational lensing of gravitational waves}",
    eprint = "2008.12814",
    archivePrefix = "arXiv",
    primaryClass = "gr-qc",
    doi = "10.1103/PhysRevD.103.064047",
    journal = "Phys. Rev. D",
    volume = "103",
    number = "6",
    pages = "064047",
    year = "2021"
}

@article{More:2021kpb,
    author = "More, Anupreeta and More, Surhud",
    title = "{Improved statistic to identify strongly lensed gravitational wave events}",
    eprint = "2111.03091",
    archivePrefix = "arXiv",
    primaryClass = "astro-ph.CO",
    doi = "10.1093/mnras/stac1704",
    journal = "Mon. Not. Roy. Astron. Soc.",
    volume = "515",
    number = "1",
    pages = "1044--1051",
    year = "2022"
}

@article{Vijaykumar:2022dlp,
    author = "Vijaykumar, Aditya and Mehta, Ajit Kumar and Ganguly, Apratim",
    title = "{Detection and parameter estimation challenges of type-II lensed binary black hole signals}",
    eprint = "2202.06334",
    archivePrefix = "arXiv",
    primaryClass = "gr-qc",
    doi = "10.1103/PhysRevD.108.043036",
    journal = "Phys. Rev. D",
    volume = "108",
    number = "4",
    pages = "043036",
    year = "2023"
}

@article{Caliskan:2022wbh,
    author = "\c{C}al\i{}\c{s}kan, Mesut and Ezquiaga, Jose Mar\'\i{}a and Hannuksela, Otto A. and Holz, Daniel E.",
    title = "{Lensing or luck? False alarm probabilities for gravitational lensing of gravitational waves}",
    eprint = "2201.04619",
    archivePrefix = "arXiv",
    primaryClass = "astro-ph.CO",
    doi = "10.1103/PhysRevD.107.063023",
    journal = "Phys. Rev. D",
    volume = "107",
    number = "6",
    pages = "063023",
    year = "2023"
}

@article{Barsode:2024zwv,
    author = "Barsode, Ankur and Goyal, Srashti and Ajith, Parameswaran",
    title = "{Fast and Efficient Bayesian Method to Search for Strongly Lensed Gravitational Waves}",
    eprint = "2412.01278",
    archivePrefix = "arXiv",
    primaryClass = "gr-qc",
    doi = "10.3847/1538-4357/adae10",
    journal = "Astrophys. J.",
    volume = "980",
    number = "2",
    pages = "258",
    year = "2025"
}

@ARTICLE{1986ApJ...307...30D,
       author = "Deguchi, S. and Watson, W.~D.",
       title = "{Diffraction in Gravitational Lensing for Compact Objects of Low Mass}",
      journal = {\apj},
         year = 1986,
        month = aug,
       volume = {307},
        pages = {30},
          doi = {10.1086/164389},
       adsurl = {https://ui.adsabs.harvard.edu/abs/1986ApJ...307...30D}
}

@article{Nakamura:1997sw,
    author = "Nakamura, Takahiro T.",
    title = "{Gravitational lensing of gravitational waves from inspiraling binaries by a point mass lens}",
    reportNumber = "UTAP-272-97, YITP-97-61",
    doi = "10.1103/PhysRevLett.80.1138",
    journal = "Phys. Rev. Lett.",
    volume = "80",
    pages = "1138--1141",
    year = "1998"
}

@article{Nakamura:1999uwi,
    author = "Nakamura, Takahiro T. and Deguchi, Shuji",
    title = "{Wave Optics in Gravitational Lensing}",
    doi = "10.1143/ptps.133.137",
    journal = "Prog. Theor. Phys. Suppl.",
    volume = "133",
    pages = "137--153",
    year = "1999"
}

@ARTICLE{2003ApJ...595.1039T,
       author = "Takahashi, Ryuichi and Nakamura, Takahiro T.",
        title = "{Wave Effects in the Gravitational Lensing of Gravitational Waves from Chirping Binaries}",
      journal = {\apj},
         year = 2003,
        month = oct,
       volume = {595},
       number = {2},
        pages = {1039-1051},
          doi = {10.1086/377430},
archivePrefix = {arXiv},
       eprint = {astro-ph/0305055},
 primaryClass = {astro-ph},
       adsurl = {https://ui.adsabs.harvard.edu/abs/2003ApJ...595.1039T}
}

@article{Jung:2017flg,
    author = "Jung, Sunghoon and Shin, Chang Sub",
    title = "{Gravitational-Wave Fringes at LIGO: Detecting Compact Dark Matter by Gravitational Lensing}",
    eprint = "1712.01396",
    archivePrefix = "arXiv",
    primaryClass = "astro-ph.CO",
    doi = "10.1103/PhysRevLett.122.041103",
    journal = "Phys. Rev. Lett.",
    volume = "122",
    number = "4",
    pages = "041103",
    year = "2019"
}

@article{Cremonese:2021ahz,
    author = "Cremonese, Paolo and Mota, David Fonseca and Salzano, Vincenzo",
    title = "{Characteristic Features of Gravitational Wave Lensing as Probe of Lens Mass Model}",
    eprint = "2111.01163",
    archivePrefix = "arXiv",
    primaryClass = "astro-ph.CO",
    doi = "10.1002/andp.202300040",
    journal = "Annalen Phys.",
    volume = "535",
    number = "6",
    pages = "2300040",
    year = "2023"
}

@article{Cremonese:2021puh,
    author = "Cremonese, Paolo and Ezquiaga, Jose Mar\'\i{}a and Salzano, Vincenzo",
    title = "{Breaking the mass-sheet degeneracy with gravitational wave interference in lensed events}",
    eprint = "2104.07055",
    archivePrefix = "arXiv",
    primaryClass = "astro-ph.CO",
    doi = "10.1103/PhysRevD.104.023503",
    journal = "Phys. Rev. D",
    volume = "104",
    number = "2",
    pages = "023503",
    year = "2021"
}

@article{Mishra:2021xzz,
    author = "Mishra, Anuj and Meena, Ashish Kumar and More, Anupreeta and Bose, Sukanta and Bagla, Jasjeet Singh",
    title = "{Gravitational lensing of gravitational waves: effect of microlens population in lensing galaxies}",
    eprint = "2102.03946",
    archivePrefix = "arXiv",
    primaryClass = "astro-ph.CO",
    doi = "10.1093/mnras/stab2875",
    journal = "Mon. Not. Roy. Astron. Soc.",
    volume = "508",
    number = "4",
    pages = "4869--4886",
    year = "2021"
}

@article{Shan:2020esq,
    author = "Shan, Xikai and Wei, Chengliang and Hu, Bin",
    title = "{Lensing magnification: gravitational waves from coalescing stellar-mass binary black holes}",
    eprint = "2012.08381",
    archivePrefix = "arXiv",
    primaryClass = "astro-ph.CO",
    doi = "10.1093/mnras/stab2567",
    journal = "Mon. Not. Roy. Astron. Soc.",
    volume = "508",
    number = "1",
    pages = "1253--1261",
    year = "2021"
}

@article{Shan:2023ngi,
    author = "Shan, Xikai and Chen, Xuechun and Hu, Bin and Cai, Rong-Gen",
    title = "{Microlensing sheds light on the detection of strong lensing gravitational waves}",
    journal = "arXiv:2301.06117",
    eprint = "2301.06117",
    archivePrefix = "arXiv",
    primaryClass = "astro-ph.IM",
    month = "1",
    year = "2023"
}

@article{Meena:2022unp,
    author = "Meena, Ashish Kumar and Mishra, Anuj and More, Anupreeta and Bose, Sukanta and Bagla, Jasjeet Singh",
    title = "{Gravitational lensing of gravitational waves: Probability of microlensing in galaxy-scale lens population}",
    eprint = "2205.05409",
    archivePrefix = "arXiv",
    primaryClass = "astro-ph.GA",
    doi = "10.1093/mnras/stac2721",
    journal = "Mon. Not. Roy. Astron. Soc.",
    volume = "517",
    number = "1",
    pages = "872--884",
    year = "2022"
}

@article{Bondarescu:2022srx,
    author = "Bondarescu, Ruxandra and Ubach, Helena and Bulashenko, Oleg and Lundgren, Andrew P.",
    title = "{Compact binaries through a lens: Silent versus detectable microlensing for the LIGO-Virgo-KAGRA gravitational wave observatories}",
    eprint = "2211.13604",
    archivePrefix = "arXiv",
    primaryClass = "gr-qc",
    doi = "10.1103/PhysRevD.108.084033",
    journal = "Phys. Rev. D",
    volume = "108",
    number = "8",
    pages = "084033",
    year = "2023"
}

@article{Mishra:2023ddt,
    author = "Mishra, Anuj and Meena, Ashish Kumar and More, Anupreeta and Bose, Sukanta",
    title = "{Exploring the impact of microlensing on gravitational wave signals: Biases, population characteristics, and prospects for detection}",
    eprint = "2306.11479",
    archivePrefix = "arXiv",
    primaryClass = "astro-ph.CO",
    doi = "10.1093/mnras/stae836",
    journal = "Mon. Not. Roy. Astron. Soc.",
    volume = "531",
    number = "1",
    pages = "764--787",
    year = "2024"
}

@article{Mishra:2023vzo,
    author = "Mishra, Anuj and Krishnendu, N. V. and Ganguly, Apratim",
    title = "{Unveiling microlensing biases in testing general relativity with gravitational waves}",
    eprint = "2311.08446",
    archivePrefix = "arXiv",
    primaryClass = "gr-qc",
    doi = "10.1103/PhysRevD.110.084009",
    journal = "Phys. Rev. D",
    volume = "110",
    number = "8",
    pages = "084009",
    year = "2024"
}

@article{Rao:2025poe,
    author = "Rao, Nishkal and Mishra, Anuj and Ganguly, Apratim and More, Anupreeta",
    title = "{Comprehensive analysis of time-domain overlapping gravitational wave transients: A Lensing Study}",
    journal = "arXiv:2510.17787",
    eprint = "2510.17787",
    archivePrefix = "arXiv",
    primaryClass = "gr-qc",
    reportNumber = "LIGO-P2500640",
    month = "10",
    year = "2025"
}

@article{LIGOScientific:2019dag,
    author = "Abbott, B. P. and others",
    collaboration = "LIGO Scientific, Virgo",
    title = "{Search for Eccentric Binary Black Hole Mergers with Advanced LIGO and Advanced Virgo during their First and Second Observing Runs}",
    eprint = "1907.09384",
    archivePrefix = "arXiv",
    primaryClass = "astro-ph.HE",
    reportNumber = "LIGO Document P1900110",
    doi = "10.3847/1538-4357/ab3c2d",
    journal = "Astrophys. J.",
    volume = "883",
    number = "2",
    pages = "149",
    year = "2019"
}

@article{Lenon:2020oza,
    author = "Lenon, Amber K. and Nitz, Alexander H. and Brown, Duncan A.",
    title = "{Measuring the eccentricity of GW170817 and GW190425}",
    eprint = "2005.14146",
    archivePrefix = "arXiv",
    primaryClass = "astro-ph.HE",
    doi = "10.1093/mnras/staa2120",
    journal = "Mon. Not. Roy. Astron. Soc.",
    volume = "497",
    number = "2",
    pages = "1966--1971",
    year = "2020"
}

@article{Romero-Shaw:2022xko,
    author = "Romero-Shaw, Isobel M. and Lasky, Paul D. and Thrane, Eric",
    title = "{Four Eccentric Mergers Increase the Evidence that LIGO\textendash{}Virgo\textendash{}KAGRA\textquoteright{}s Binary Black Holes Form Dynamically}",
    eprint = "2206.14695",
    archivePrefix = "arXiv",
    primaryClass = "astro-ph.HE",
    doi = "10.3847/1538-4357/ac9798",
    journal = "Astrophys. J.",
    volume = "940",
    number = "2",
    pages = "171",
    year = "2022"
}

@article{Iglesias:2022xfc,
    author = "Iglesias, H. L. and others",
    title = "{Eccentricity Estimation for Five Binary Black Hole Mergers with Higher-order Gravitational-wave Modes}",
    eprint = "2208.01766",
    archivePrefix = "arXiv",
    primaryClass = "gr-qc",
    reportNumber = "LIGO-P2200208",
    doi = "10.3847/1538-4357/ad5ff6",
    journal = "Astrophys. J.",
    volume = "972",
    number = "1",
    pages = "65",
    year = "2024"
}

@article{Dhurkunde:2023qoe,
    author = "Dhurkunde, Rahul and Nitz, Alexander H.",
    title = "{Search for eccentric NSBH and BNS mergers in the third observing run of Advanced LIGO and Virgo}",
    eprint = "2311.00242",
    archivePrefix = "arXiv",
    primaryClass = "astro-ph.HE",
    doi = "10.1103/PhysRevD.111.103018",
    journal = "Phys. Rev. D",
    volume = "111",
    number = "10",
    pages = "103018",
    year = "2025"
}

@article{Gupte:2024jfe,
    author = "Gupte, Nihar and others",
    title = "{Evidence for eccentricity in the population of binary black holes observed by LIGO-Virgo-KAGRA}",
    eprint = "2404.14286",
    archivePrefix = "arXiv",
    primaryClass = "gr-qc",
    doi = "10.1103/vpyp-nvfp",
    journal = "Phys. Rev. D",
    volume = "112",
    number = "10",
    pages = "104045",
    year = "2025"
}

@article{McMillin:2025hof,
    author = "McMillin, Patricia and Wagner, Katelyn J. and Ficarra, Giuseppe and Lousto, Carlos O. and O'Shaughnessy, Richard",
    title = "{Parameter Estimation for GW200208{\_}22 with Targeted Eccentric Numerical-relativity Simulations}",
    journal = "arXiv:2507.22862",
    eprint = "2507.22862",
    archivePrefix = "arXiv",
    primaryClass = "gr-qc",
    month = "7",
    year = "2025"
}

@article{LIGOScientific:2025brd,
    author = "Abac, A. G. and others",
    collaboration = "LIGO Scientific, Virgo, KAGRA",
    title = "{GW241011 and GW241110: Exploring Binary Formation and Fundamental Physics with Asymmetric, High-spin Black Hole Coalescences}",
    eprint = "2510.26931",
    archivePrefix = "arXiv",
    primaryClass = "astro-ph.HE",
    reportNumber = "LIGO-P2500402",
    doi = "10.3847/2041-8213/ae0d54",
    journal = "Astrophys. J. Lett.",
    volume = "993",
    number = "1",
    pages = "L21",
    year = "2025"
}

@article{Tiwari:2025fua,
    author = "Tiwari, Avinash and Bhat, Sajad A. and Shaikh, Md Arif and Kapaida, Shashvath J.",
    title = "{Testing the nature of GW200105 by probing the frequency evolution of eccentricity}",
    journal = "arXiv:2509.26152",
    eprint = "2509.26152",
    archivePrefix = "arXiv",
    primaryClass = "astro-ph.HE",
    month = "9",
    year = "2025"
}

@article{Romero-Shaw:2025vbc,
    author = "Romero-Shaw, Isobel and Stegmann, Jakob and Tagawa, Hiromichi and Gerosa, Davide and Samsing, Johan and Gupte, Nihar and Green, Stephen R.",
    title = "{GW200208{\_}222617 as an eccentric black-hole binary merger: Properties and astrophysical implications}",
    eprint = "2506.17105",
    archivePrefix = "arXiv",
    primaryClass = "astro-ph.HE",
    doi = "10.1103/jj7m-x66y",
    journal = "Phys. Rev. D",
    volume = "112",
    number = "6",
    pages = "063052",
    year = "2025"
}

@article{Kacanja:2025kpr,
    author = "Kacanja, Keisi and Soni, Kanchan and Nitz, Alexander Harvey",
    title = "{Eccentricity signatures in LIGO-Virgo-KAGRA's BNS and NSBH binaries}",
    journal = "arXiv:2508.00179",
    eprint = "2508.00179",
    archivePrefix = "arXiv",
    primaryClass = "gr-qc",
    month = "7",
    year = "2025"
}

@article{Bethe:1998bn,
    author = "Bethe, Hans A. and Brown, G. E.",
    title = "{Evolution of binary compact objects which merge}",
    eprint = "astro-ph/9802084",
    archivePrefix = "arXiv",
    reportNumber = "SUNY-NTG-98-4",
    doi = "10.1086/306265",
    journal = "Astrophys. J.",
    volume = "506",
    pages = "780--789",
    year = "1998"
}

@article{Ng:2020qpk,
    author = "Ng, Ken K. Y. and Vitale, Salvatore and Farr, Will M. and Rodriguez, Carl L.",
    title = "{Probing multiple populations of compact binaries with third-generation gravitational-wave detectors}",
    eprint = "2012.09876",
    archivePrefix = "arXiv",
    primaryClass = "astro-ph.CO",
    reportNumber = "LIGO-P2000540",
    doi = "10.3847/2041-8213/abf8be",
    journal = "Astrophys. J. Lett.",
    volume = "913",
    number = "1",
    pages = "L5",
    year = "2021"
}

@article{Wang:2023tle,
    author = "Wang, Han and Harry, Ian and Nitz, Alexander and Hu, Yi-Ming",
    title = "{Space-based gravitational wave observatories will be able to use eccentricity to unveil stellar-mass binary black hole formation}",
    eprint = "2304.10340",
    archivePrefix = "arXiv",
    primaryClass = "astro-ph.HE",
    doi = "10.1103/PhysRevD.109.063029",
    journal = "Phys. Rev. D",
    volume = "109",
    number = "6",
    pages = "063029",
    year = "2024"
}

@article{Yang:2024vfy,
    author = "Yang, Tao and Cai, Rong-Gen and Cao, Zhoujian and Lee, Hyung Mok",
    title = "{The Advantage of Early Detection and Localization from Eccentricity-Induced Higher Harmonic Modes in Second-Generation Ground-Based Detector Networks}",
    journal = "arXiv:2412.20664",
    eprint = "2412.20664",
    archivePrefix = "arXiv",
    primaryClass = "gr-qc",
    month = "12",
    year = "2024"
}

@article{Dai:2018enj,
    author = "Dai, Liang and Li, Shun-Sheng and Zackay, Barak and Mao, Shude and Lu, Youjun",
    title = "{Detecting Lensing-Induced Diffraction in Astrophysical Gravitational Waves}",
    eprint = "1810.00003",
    archivePrefix = "arXiv",
    primaryClass = "gr-qc",
    doi = "10.1103/PhysRevD.98.104029",
    journal = "Phys. Rev. D",
    volume = "98",
    number = "10",
    pages = "104029",
    year = "2018"
}

@article{Hinder:2017sxy,
    author = "Hinder, Ian and Kidder, Lawrence E. and Pfeiffer, Harald P.",
    title = "{Eccentric binary black hole inspiral-merger-ringdown gravitational waveform model from numerical relativity and post-Newtonian theory}",
    eprint = "1709.02007",
    archivePrefix = "arXiv",
    primaryClass = "gr-qc",
    doi = "10.1103/PhysRevD.98.044015",
    journal = "Phys. Rev. D",
    volume = "98",
    number = "4",
    pages = "044015",
    year = "2018"
}

@article{Liu:2020par,
    author = "Liu, Xiaoshu and Magana Hernandez, Ignacio and Creighton, Jolien",
    title = "{Identifying strong gravitational-wave lensing during the second observing run of Advanced LIGO and Advanced Virgo}",
    eprint = "2009.06539",
    archivePrefix = "arXiv",
    primaryClass = "astro-ph.HE",
    doi = "10.3847/1538-4357/abd7eb",
    journal = "Astrophys. J.",
    volume = "908",
    number = "1",
    pages = "97",
    year = "2021"
}

@article{Janquart:2023mvf,
    author = "Janquart, Justin and others",
    title = "{Follow-up analyses to the O3 LIGO{\textendash}Virgo{\textendash}KAGRA lensing searches}",
    eprint = "2306.03827",
    archivePrefix = "arXiv",
    primaryClass = "gr-qc",
    doi = "10.1093/mnras/stad2909",
    journal = "Mon. Not. Roy. Astron. Soc.",
    volume = "526",
    number = "3",
    pages = "3832--3860",
    year = "2023"
}

@article{LIGOScientific:2023bwz,
    author = "Abbott, R. and others",
    collaboration = "LIGO Scientific, KAGRA, VIRGO",
    title = "{Search for Gravitational-lensing Signatures in the Full Third Observing Run of the LIGO{\textendash}Virgo Network}",
    eprint = "2304.08393",
    archivePrefix = "arXiv",
    primaryClass = "gr-qc",
    reportNumber = "LIGO-P2200031",
    doi = "10.3847/1538-4357/ad3e83",
    journal = "Astrophys. J.",
    volume = "970",
    number = "2",
    pages = "191",
    year = "2024"
}

@article{Goyal:2020bkm,
    author = "Goyal, Srashti and Haris, K. and Mehta, Ajit Kumar and Ajith, Parameswaran",
    title = "{Testing the nature of gravitational-wave polarizations using strongly lensed signals}",
    eprint = "2008.07060",
    archivePrefix = "arXiv",
    primaryClass = "gr-qc",
    reportNumber = "LIGO-P2000295-v1",
    doi = "10.1103/PhysRevD.103.024038",
    journal = "Phys. Rev. D",
    volume = "103",
    number = "2",
    pages = "024038",
    year = "2021"
}

@article{Mukherjee:2019wcg,
    author = "Mukherjee, Suvodip and Wandelt, Benjamin D. and Silk, Joseph",
    title = "{Probing the theory of gravity with gravitational lensing of gravitational waves and galaxy surveys}",
    eprint = "1908.08951",
    archivePrefix = "arXiv",
    primaryClass = "astro-ph.CO",
    doi = "10.1093/mnras/staa827",
    journal = "Mon. Not. Roy. Astron. Soc.",
    volume = "494",
    number = "2",
    pages = "1956--1970",
    year = "2020"
}

@article{Hannuksela:2019kle,
    author = "Hannuksela, O. A. and Haris, K. and Ng, K. K. Y. and Kumar, S. and Mehta, A. K. and Keitel, D. and Li, T. G. F. and Ajith, P.",
    title = "{Search for gravitational lensing signatures in LIGO-Virgo binary black hole events}",
    eprint = "1901.02674",
    archivePrefix = "arXiv",
    primaryClass = "gr-qc",
    reportNumber = "LIGO Document P1800297, LIGO-P1800297",
    doi = "10.3847/2041-8213/ab0c0f",
    journal = "Astrophys. J. Lett.",
    volume = "874",
    number = "1",
    pages = "L2",
    year = "2019"
}

@article{Jana:2022shb,
    author = "Jana, Souvik and Kapadia, Shasvath J. and Venumadhav, Tejaswi and Ajith, Parameswaran",
    title = "{Cosmography Using Strongly Lensed Gravitational Waves from Binary Black Holes}",
    eprint = "2211.12212",
    archivePrefix = "arXiv",
    primaryClass = "astro-ph.CO",
    reportNumber = "LIGO-P2200298",
    doi = "10.1103/PhysRevLett.130.261401",
    journal = "Phys. Rev. Lett.",
    volume = "130",
    number = "26",
    pages = "261401",
    year = "2023"
}

@article{Hannuksela:2020xor,
    author = "Hannuksela, Otto A. and Collett, Thomas E. and {\c{C}}al{\i}{\c{s}}kan, Mesut and Li, Tjonnie G. F.",
    title = "{Localizing merging black holes with sub-arcsecond precision using gravitational-wave lensing}",
    eprint = "2004.13811",
    archivePrefix = "arXiv",
    primaryClass = "astro-ph.HE",
    doi = "10.1093/mnras/staa2577",
    journal = "Mon. Not. Roy. Astron. Soc.",
    volume = "498",
    number = "3",
    pages = "3395--3402",
    year = "2020"
}

@article{Liao:2017ioi,
    author = "Liao, Kai and Fan, Xi-Long and Ding, Xu-Heng and Biesiada, Marek and Zhu, Zong-Hong",
    title = "{Precision cosmology from future lensed gravitational wave and electromagnetic signals}",
    eprint = "1703.04151",
    archivePrefix = "arXiv",
    primaryClass = "astro-ph.CO",
    doi = "10.1038/s41467-017-01152-9",
    journal = "Nature Commun.",
    volume = "8",
    number = "1",
    pages = "1148",
    year = "2017",
    note = "[Erratum: Nature Commun. 8, 2136 (2017)]"
}

@article{Baker:2016reh,
    author = "Baker, Tessa and Trodden, Mark",
    title = "{Multimessenger time delays from lensed gravitational waves}",
    eprint = "1612.02004",
    archivePrefix = "arXiv",
    primaryClass = "astro-ph.CO",
    doi = "10.1103/PhysRevD.95.063512",
    journal = "Phys. Rev. D",
    volume = "95",
    number = "6",
    pages = "063512",
    year = "2017"
}

@article{Mukherjee:2021qam,
    author = "Mukherjee, Suvodip and Broadhurst, Tom and Diego, Jose M. and Silk, Joseph and Smoot, George F.",
    title = "{Impact of astrophysical binary coalescence time-scales on the rate of lensed gravitational wave events}",
    eprint = "2106.00392",
    archivePrefix = "arXiv",
    primaryClass = "gr-qc",
    doi = "10.1093/mnras/stab1980",
    journal = "Mon. Not. Roy. Astron. Soc.",
    volume = "506",
    number = "3",
    pages = "3751--3759",
    year = "2021"
}

@article{Basak:2021ten,
    author = "Basak, S. and Ganguly, A. and Haris, K. and Kapadia, S. and Mehta, A. K. and Ajith, P.",
    title = "{Constraints on Compact Dark Matter from Gravitational Wave Microlensing}",
    eprint = "2109.06456",
    archivePrefix = "arXiv",
    primaryClass = "gr-qc",
    reportNumber = "LIGO-P2100321",
    doi = "10.3847/2041-8213/ac4dfa",
    journal = "Astrophys. J.",
    volume = "926",
    number = "2",
    pages = "L28",
    year = "2022"
}

@article{Ajith:2012mn,
    author = "Ajith, P. and Fotopoulos, N. and Privitera, S. and Neunzert, A. and Weinstein, A. J.",
    title = "{Effectual template bank for the detection of gravitational waves from inspiralling compact binaries with generic spins}",
    eprint = "1210.6666",
    archivePrefix = "arXiv",
    primaryClass = "gr-qc",
    reportNumber = "LIGO-P1200106-V2, LIGO-P1200106-V3",
    doi = "10.1103/PhysRevD.89.084041",
    journal = "Phys. Rev. D",
    volume = "89",
    number = "8",
    pages = "084041",
    year = "2014"
}

@article{Peters:1964zz,
    author = "Peters, P. C.",
    title = "{Gravitational Radiation and the Motion of Two Point Masses}",
    doi = "10.1103/PhysRev.136.B1224",
    journal = "Phys. Rev.",
    volume = "136",
    pages = "B1224--B1232",
    year = "1964"
}

@article{Zevin:2018kzq,
    author = "Zevin, Michael and Samsing, Johan and Rodriguez, Carl and Haster, Carl-Johan and Ramirez-Ruiz, Enrico",
    title = "{Eccentric Black Hole Mergers in Dense Star Clusters: The Role of Binary\textendash{}Binary Encounters}",
    eprint = "1810.00901",
    archivePrefix = "arXiv",
    primaryClass = "astro-ph.HE",
    reportNumber = "LIGO-P1800275",
    doi = "10.3847/1538-4357/aaf6ec",
    journal = "Astrophys. J.",
    volume = "871",
    number = "1",
    pages = "91",
    year = "2019"
}

@inbook{Mapelli:2021taw,
    author = "Mapelli, Michela",
    title = "{Formation Channels of Single and Binary Stellar-Mass Black Holes}",
    eprint = "2106.00699",
    archivePrefix = "arXiv",
    primaryClass = "astro-ph.HE",
    doi = "10.1007/978-981-15-4702-7_16-1",
    year = "2021"
}

@article{Cholis:2016kqi,
    author = {Cholis, Ilias and Kovetz, Ely D. and Ali-Ha{\"\i}moud, Yacine and Bird, Simeon and Kamionkowski, Marc and Mu{\~n}oz, Julian B. and Raccanelli, Alvise},
    title = "{Orbital eccentricities in primordial black hole binaries}",
    eprint = "1606.07437",
    archivePrefix = "arXiv",
    primaryClass = "astro-ph.HE",
    doi = "10.1103/PhysRevD.94.084013",
    journal = "Phys. Rev. D",
    volume = "94",
    number = "8",
    pages = "084013",
    year = "2016"
}

@article{Wen:2002km,
    author = "Wen, Linqing",
    title = "{On the eccentricity distribution of coalescing black hole binaries driven by the Kozai mechanism in globular clusters}",
    eprint = "astro-ph/0211492",
    archivePrefix = "arXiv",
    doi = "10.1086/378794",
    journal = "Astrophys. J.",
    volume = "598",
    pages = "419--430",
    year = "2003"
}

@article{Antonini:2013tea,
    author = "Antonini, Fabio and Murray, Norman and Mikkola, Seppo",
    title = "{Black hole triple dynamics: breakdown of the orbit average approximation and implications for gravitational wave detections}",
    eprint = "1308.3674",
    archivePrefix = "arXiv",
    primaryClass = "astro-ph.HE",
    doi = "10.1088/0004-637X/781/1/45",
    journal = "Astrophys. J.",
    volume = "781",
    pages = "45",
    year = "2014"
}

@article{Lai:2018rto,
    author = "Lai, Kwun-Hang and Hannuksela, Otto A. and Herrera-Mart{\'\i}n, Antonio and Diego, Jose M. and Broadhurst, Tom and Li, Tjonnie G. F.",
    title = "{Discovering intermediate-mass black hole lenses through gravitational wave lensing}",
    eprint = "1801.07840",
    archivePrefix = "arXiv",
    primaryClass = "gr-qc",
    doi = "10.1103/PhysRevD.98.083005",
    journal = "Phys. Rev. D",
    volume = "98",
    number = "8",
    pages = "083005",
    year = "2018"
}

@article{OShea:2021faf,
    author = "O'Shea, Eamonn and Kumar, Prayush",
    title = "{Correlations in gravitational-wave reconstructions from eccentric binaries: A case study with GW151226 and GW170608}",
    eprint = "2107.07981",
    archivePrefix = "arXiv",
    primaryClass = "astro-ph.HE",
    doi = "10.1103/PhysRevD.108.104018",
    journal = "Phys. Rev. D",
    volume = "108",
    number = "10",
    pages = "104018",
    year = "2023"
}

@article{Shaikh:2023ypz,
    author = "Shaikh, Md Arif and Varma, Vijay and Pfeiffer, Harald P. and Ramos-Buades, Antoni and van de Meent, Maarten",
    title = "{Defining eccentricity for gravitational wave astronomy}",
    eprint = "2302.11257",
    archivePrefix = "arXiv",
    primaryClass = "gr-qc",
    doi = "10.1103/PhysRevD.108.104007",
    journal = "Phys. Rev. D",
    volume = "108",
    number = "10",
    pages = "104007",
    year = "2023"
}

@article{Shaikh:2025tae,
    author = "Shaikh, Md Arif and Varma, Vijay and Ramos-Buades, Antoni and Pfeiffer, Harald P. and Boyle, Michael and Kidder, Lawrence E. and Scheel, Mark A.",
    title = "{Defining eccentricity for spin-precessing binaries}",
    eprint = "2507.08345",
    archivePrefix = "arXiv",
    primaryClass = "gr-qc",
    doi = "10.1088/1361-6382/ae085d",
    journal = "Class. Quant. Grav.",
    volume = "42",
    number = "19",
    pages = "195012",
    year = "2025"
}

@article{Scheel:2025jct,
    author = "Scheel, Mark A. and others",
    title = "{The SXS collaboration{\textquoteright}s third catalog of binary black hole simulations}",
    eprint = "2505.13378",
    archivePrefix = "arXiv",
    primaryClass = "gr-qc",
    doi = "10.1088/1361-6382/adfd34",
    journal = "Class. Quant. Grav.",
    volume = "42",
    number = "19",
    pages = "195017",
    year = "2025"
}

@article{LIGOScientific:2016aoc,
    author = "Abbott, B. P. and others",
    collaboration = "LIGO Scientific, Virgo",
    title = "{Observation of Gravitational Waves from a Binary Black Hole Merger}",
    eprint = "1602.03837",
    archivePrefix = "arXiv",
    primaryClass = "gr-qc",
    reportNumber = "LIGO-P150914",
    doi = "10.1103/PhysRevLett.116.061102",
    journal = "Phys. Rev. Lett.",
    volume = "116",
    number = "6",
    pages = "061102",
    year = "2016"
}

@article{Colleoni:2024knd,
    author = "Colleoni, Marta and Vidal, Felip A. Ramis and Garc{\'\i}a-Quir{\'o}s, Cecilio and Ak{\c{c}}ay, Sarp and Bera, Sayantani",
    title = "{Fast frequency-domain gravitational waveforms for precessing binaries with a new twist}",
    eprint = "2412.16721",
    archivePrefix = "arXiv",
    primaryClass = "gr-qc",
    doi = "10.1103/PhysRevD.111.104019",
    journal = "Phys. Rev. D",
    volume = "111",
    number = "10",
    pages = "104019",
    year = "2025"
}

@article{Pratten:2020ceb,
    author = "Pratten, Geraint and others",
    title = "{Computationally efficient models for the dominant and subdominant harmonic modes of precessing binary black holes}",
    eprint = "2004.06503",
    archivePrefix = "arXiv",
    primaryClass = "gr-qc",
    doi = "10.1103/PhysRevD.103.104056",
    journal = "Phys. Rev. D",
    volume = "103",
    number = "10",
    pages = "104056",
    year = "2021"
}

@article{behnel2011cython,
title={Cython: The best of both worlds},
author={Behnel, Stefan and Bradshaw, Robert and Citro, Craig and Dalcin, Lisandro and Seljebotn, Dag Sverre and Smith, Kurt},
journal={Computing in Science \& Engineering},
volume={13},
number={2},
pages={31--39},
year={2011},
publisher={IEEE}
}

@article{Harris:2020xlr,
    author = "Harris, Charles R. and others",
    title = "{Array programming with NumPy}",
    eprint = "2006.10256",
    archivePrefix = "arXiv",
    primaryClass = "cs.MS",
    doi = "10.1038/s41586-020-2649-2",
    journal = "Nature",
    volume = "585",
    number = "7825",
    pages = "357--362",
    year = "2020"
}

@Article{Hunter:2007,
  Author    = {Hunter, J. D.},
  Title     = {Matplotlib: A 2D graphics environment},
  Journal   = {Computing in Science \& Engineering},
  Volume    = {9},
  Number    = {3},
  Pages     = {90--95},
  publisher = {IEEE COMPUTER SOC},
  doi       = {10.1109/MCSE.2007.55},
  year      = 2007
}

@conference{Kluyver2016jupyter,
    Title = {Jupyter Notebooks -- a publishing format for reproducible computational workflows},
    Author = {Thomas Kluyver and others},
    Booktitle = {Positioning and Power in Academic Publishing: Players, Agents and Agendas},
    Editor = {F. Loizides and B. Schmidt},
    Organization = {IOS Press},
    Pages = {87 - 90},
    Year = {2016}
}

@ARTICLE{2020MNRAS.493.3132S,
       author = {{Speagle}, Joshua S.},
        title = "{DYNESTY: a dynamic nested sampling package for estimating Bayesian posteriors and evidences}",
      journal = {MNRAS},
         year = 2020,
        month = apr,
       volume = {493},
       number = {3},
        pages = {3132-3158},
          doi = {10.1093/mnras/staa278},
archivePrefix = {arXiv},
       eprint = {1904.02180},
 primaryClass = {astro-ph.IM},
       adsurl = {https://ui.adsabs.harvard.edu/abs/2020MNRAS.493.3132S}
}

@article{Hoy:2020vys,
    author = "Hoy, Charlie and Raymond, Vivien",
    title = "{PESummary: the code agnostic Parameter Estimation Summary page builder}",
    eprint = "2006.06639",
    archivePrefix = "arXiv",
    primaryClass = "astro-ph.IM",
    reportNumber = "LIGO-P2000156",
    doi = "10.1016/j.softx.2021.100765",
    journal = "SoftwareX",
    volume = "15",
    pages = "100765",
    year = "2021"
}

@ARTICLE{2019PhRvD.100d3030T,
       author = {{Talbot}, Colm and {Smith}, Rory and {Thrane}, Eric and
         {Poole}, Gregory B.},
        title = "{Parallelized inference for gravitational-wave astronomy}",
      journal = {\prd},
         year = 2019,
        month = aug,
       volume = {100},
       number = {4},
          eid = {043030},
        pages = {043030},
          doi = {10.1103/PhysRevD.100.043030},
archivePrefix = {arXiv},
       eprint = {1904.02863},
 primaryClass = {astro-ph.IM},
}

@article{Virtanen:2019joe,
    author = "Virtanen, Pauli and others",
    title = "{SciPy 1.0--Fundamental Algorithms for Scientific Computing in Python}",
    eprint = "1907.10121",
    archivePrefix = "arXiv",
    primaryClass = "cs.MS",
    doi = "10.1038/s41592-019-0686-2",
    journal = "Nature Meth.",
    volume = "17",
    pages = "261",
    year = "2020"
}

@MISC{2020ascl.soft12021L,
       author = {{LIGO Scientific Collaboration}},
        title = "{LALSuite: LIGO Scientific Collaboration Algorithm Library Suite}",
 howpublished = {Astrophysics Source Code Library, record ascl:2012.021},
         year = 2020,
        month = dec,
          eid = {ascl:2012.021},
        pages = {ascl:2012.021},
archivePrefix = {ascl},
       eprint = {2012.021},
       adsurl = {https://ui.adsabs.harvard.edu/abs/2020ascl.soft12021L}
}

@misc{pycbc_github,
  author = {Alex Nitz and others},
  title = {PyCBC Software},
  year = {2024},
  doi  = {10.5281/zenodo.596388},
  howpublished = {\url{https://github.com/gwastro/pycbc}},
}

@article{Damour:2014sva,
    author = "Damour, Thibault and Nagar, Alessandro",
    title = "{New effective-one-body description of coalescing nonprecessing spinning black-hole binaries}",
    eprint = "1406.6913",
    archivePrefix = "arXiv",
    primaryClass = "gr-qc",
    doi = "10.1103/PhysRevD.90.044018",
    journal = "Phys. Rev. D",
    volume = "90",
    number = "4",
    pages = "044018",
    year = "2014"
}

@article{Nagar:2015xqa,
    author = "Nagar, Alessandro and Damour, Thibault and Reisswig, Christian and Pollney, Denis",
    title = "{Energetics and phasing of nonprecessing spinning coalescing black hole binaries}",
    eprint = "1506.08457",
    archivePrefix = "arXiv",
    primaryClass = "gr-qc",
    doi = "10.1103/PhysRevD.93.044046",
    journal = "Phys. Rev. D",
    volume = "93",
    number = "4",
    pages = "044046",
    year = "2016"
}

@article{Nagar:2018zoe,
    author = "Nagar, Alessandro and others",
    title = "{Time-domain effective-one-body gravitational waveforms for coalescing compact binaries with nonprecessing spins, tides and self-spin effects}",
    eprint = "1806.01772",
    archivePrefix = "arXiv",
    primaryClass = "gr-qc",
    doi = "10.1103/PhysRevD.98.104052",
    journal = "Phys. Rev. D",
    volume = "98",
    number = "10",
    pages = "104052",
    year = "2018"
}

@article{Nagar:2019wds,
    author = "Nagar, Alessandro and Pratten, Geraint and Riemenschneider, Gunnar and Gamba, Rossella",
    title = "{Multipolar effective one body model for nonspinning black hole binaries}",
    eprint = "1904.09550",
    archivePrefix = "arXiv",
    primaryClass = "gr-qc",
    doi = "10.1103/PhysRevD.101.024041",
    journal = "Phys. Rev. D",
    volume = "101",
    number = "2",
    pages = "024041",
    year = "2020"
}

@article{Nagar:2020pcj,
    author = "Nagar, Alessandro and Riemenschneider, Gunnar and Pratten, Geraint and Rettegno, Piero and Messina, Francesco",
    title = "{Multipolar effective one body waveform model for spin-aligned black hole binaries}",
    eprint = "2001.09082",
    archivePrefix = "arXiv",
    primaryClass = "gr-qc",
    doi = "10.1103/PhysRevD.102.024077",
    journal = "Phys. Rev. D",
    volume = "102",
    number = "2",
    pages = "024077",
    year = "2020"
}

@article{Riemenschneider:2021ppj,
    author = "Riemenschneider, Gunnar and Rettegno, Piero and Breschi, Matteo and Albertini, Angelica and Gamba, Rossella and Bernuzzi, Sebastiano and Nagar, Alessandro",
    title = "{Assessment of consistent next-to-quasicircular corrections and postadiabatic approximation in effective-one-body multipolar waveforms for binary black hole coalescences}",
    eprint = "2104.07533",
    archivePrefix = "arXiv",
    primaryClass = "gr-qc",
    doi = "10.1103/PhysRevD.104.104045",
    journal = "Phys. Rev. D",
    volume = "104",
    number = "10",
    pages = "104045",
    year = "2021"
}

@article{Nagar:2023zxh,
    author = "Nagar, Alessandro and Rettegno, Piero and Gamba, Rossella and Albanesi, Simone and Albertini, Angelica and Bernuzzi, Sebastiano",
    title = "{Analytic systematics in next generation of effective-one-body gravitational waveform models for future observations}",
    eprint = "2304.09662",
    archivePrefix = "arXiv",
    primaryClass = "gr-qc",
    doi = "10.1103/PhysRevD.108.124018",
    journal = "Phys. Rev. D",
    volume = "108",
    number = "12",
    pages = "124018",
    year = "2023"
}

@article{Khan:2018fmp,
    author = "Khan, Sebastian and Chatziioannou, Katerina and Hannam, Mark and Ohme, Frank",
    title = "{Phenomenological model for the gravitational-wave signal from precessing binary black holes with two-spin effects}",
    eprint = "1809.10113",
    archivePrefix = "arXiv",
    primaryClass = "gr-qc",
    doi = "10.1103/PhysRevD.100.024059",
    journal = "Phys. Rev. D",
    volume = "100",
    number = "2",
    pages = "024059",
    year = "2019"
}

@article{Garcia-Quiros:2020qpx,
    author = "Garc{\'\i}a-Quir{\'o}s, Cecilio and Colleoni, Marta and Husa, Sascha and Estell{\'e}s, H{\'e}ctor and Pratten, Geraint and Ramos-Buades, Antoni and Mateu-Lucena, Maite and Jaume, Rafel",
    title = "{Multimode frequency-domain model for the gravitational wave signal from nonprecessing black-hole binaries}",
    eprint = "2001.10914",
    archivePrefix = "arXiv",
    primaryClass = "gr-qc",
    doi = "10.1103/PhysRevD.102.064002",
    journal = "Phys. Rev. D",
    volume = "102",
    number = "6",
    pages = "064002",
    year = "2020"
}

@book{Jeffreys:1939xee,
    author = "Jeffreys, Harold",
    title = "{The Theory of Probability}",
    isbn = "978-0-19-850368-2, 978-0-19-853193-7",
    series = "Oxford Classic Texts in the Physical Sciences",
    year = "1939"
}

@article{Chiaramello:2020ehz,
    author = "Chiaramello, Danilo and Nagar, Alessandro",
    title = "{Faithful analytical effective-one-body waveform model for spin-aligned, moderately eccentric, coalescing black hole binaries}",
    eprint = "2001.11736",
    archivePrefix = "arXiv",
    primaryClass = "gr-qc",
    doi = "10.1103/PhysRevD.101.101501",
    journal = "Phys. Rev. D",
    volume = "101",
    number = "10",
    pages = "101501",
    year = "2020"
}

@article{Romero-Shaw:2020thy,
    author = "Romero-Shaw, Isobel M. and Lasky, Paul D. and Thrane, Eric and Bustillo, Juan Calderon",
    title = "{GW190521: orbital eccentricity and signatures of dynamical formation in a binary black hole merger signal}",
    eprint = "2009.04771",
    archivePrefix = "arXiv",
    primaryClass = "astro-ph.HE",
    doi = "10.3847/2041-8213/abbe26",
    journal = "Astrophys. J. Lett.",
    volume = "903",
    number = "1",
    pages = "L5",
    year = "2020"
}

@article{Favata:2021vhw,
    author = "Favata, Marc and Kim, Chunglee and Arun, K. G. and Kim, JeongCho and Lee, Hyung Won",
    title = "{Constraining the orbital eccentricity of inspiralling compact binary systems with Advanced LIGO}",
    eprint = "2108.05861",
    archivePrefix = "arXiv",
    primaryClass = "gr-qc",
    reportNumber = "LIGO DCC P2100284",
    doi = "10.1103/PhysRevD.105.023003",
    journal = "Phys. Rev. D",
    volume = "105",
    number = "2",
    pages = "023003",
    year = "2022"
}

@article{Romero-Shaw:2022fbf,
    author = "Romero-Shaw, Isobel M. and Gerosa, Davide and Loutrel, Nicholas",
    title = "{Eccentricity or spin precession? Distinguishing subdominant effects in gravitational-wave data}",
    eprint = "2211.07528",
    archivePrefix = "arXiv",
    primaryClass = "astro-ph.HE",
    doi = "10.1093/mnras/stad031",
    journal = "Mon. Not. Roy. Astron. Soc.",
    volume = "519",
    number = "4",
    pages = "5352--5357",
    year = "2023"
}

@article{CalderonBustillo:2020xms,
    author = "Calder{\'o}n Bustillo, Juan and Sanchis-Gual, Nicolas and Torres-Forn{\'e}, Alejandro and Font, Jos{\'e} A.",
    title = "{Confusing Head-On Collisions with Precessing Intermediate-Mass Binary Black Hole Mergers}",
    eprint = "2009.01066",
    archivePrefix = "arXiv",
    primaryClass = "gr-qc",
    reportNumber = "LIGO-P1900363",
    doi = "10.1103/PhysRevLett.126.201101",
    journal = "Phys. Rev. Lett.",
    volume = "126",
    number = "20",
    pages = "201101",
    year = "2021"
}

@article{Xu:2022zza,
    author = "Xu, Yumeng and Hamilton, Eleanor",
    title = "{Measurability of precession and eccentricity for heavy binary-black-hole mergers}",
    eprint = "2211.09561",
    archivePrefix = "arXiv",
    primaryClass = "gr-qc",
    doi = "10.1103/PhysRevD.107.103049",
    journal = "Phys. Rev. D",
    volume = "107",
    number = "10",
    pages = "103049",
    year = "2023"
}

@article{Hegde:2023yoz,
    author = "Hegde, Ravikiran and Bose, Nirban and Pai, Archana",
    title = "{Probing eccentric higher-order modes through an effective chirp-mass model}",
    eprint = "2310.13662",
    archivePrefix = "arXiv",
    primaryClass = "gr-qc",
    reportNumber = "LIGO-P2300304",
    doi = "10.1103/PhysRevD.110.044026",
    journal = "Phys. Rev. D",
    volume = "110",
    number = "4",
    pages = "044026",
    year = "2024"
}

@article{Patterson:2024vbo,
    author = "Patterson, Ben G. and Tomson, Sharon Mary and Fairhurst, Stephen",
    title = "{Identifying eccentricity in binary black hole mergers using a harmonic decomposition of the gravitational waveform}",
    eprint = "2411.04187",
    archivePrefix = "arXiv",
    primaryClass = "gr-qc",
    doi = "10.1103/PhysRevD.111.044073",
    journal = "Phys. Rev. D",
    volume = "111",
    number = "4",
    pages = "044073",
    year = "2025"
}

@article{Talbot:2019okv,
	author = "Talbot, Colm and Smith, Rory and Thrane, Eric and Poole, Gregory B.",
	title = "{Parallelized Inference for Gravitational-Wave Astronomy}",
	eprint = "1904.02863",
	archivePrefix = "arXiv",
	primaryClass = "astro-ph.IM",
	doi = "10.1103/PhysRevD.100.043030",
	journal = "Phys. Rev. D",
	volume = "100",
	number = "4",
	pages = "043030",
	year = "2019"
}

@article{Fishbach:2018edt,
	author = "Fishbach, Maya and Holz, Daniel E. and Farr, Will M.",
	title = "{Does the Black Hole Merger Rate Evolve with Redshift?}",
	eprint = "1805.10270",
	archivePrefix = "arXiv",
	primaryClass = "astro-ph.HE",
	doi = "10.3847/2041-8213/aad800",
	journal = "Astrophys. J. Lett.",
	volume = "863",
	number = "2",
	pages = "L41",
	year = "2018"
}

@article{Madau:2014bja,
    author = "Madau, Piero and Dickinson, Mark",
    title = "{Cosmic Star Formation History}",
    eprint = "1403.0007",
    archivePrefix = "arXiv",
    primaryClass = "astro-ph.CO",
    doi = "10.1146/annurev-astro-081811-125615",
    journal = "Ann. Rev. Astron. Astrophys.",
    volume = "52",
    pages = "415--486",
    year = "2014"
}

@article{Planck:2015fie,
    author = "Ade, P. A. R. and others",
    collaboration = "Planck",
    title = "{Planck 2015 results. XIII. Cosmological parameters}",
    eprint = "1502.01589",
    archivePrefix = "arXiv",
    primaryClass = "astro-ph.CO",
    doi = "10.1051/0004-6361/201525830",
    journal = "Astron. Astrophys.",
    volume = "594",
    pages = "A13",
    year = "2016"
}

@article{Nelder:1965zz,
    author = "Nelder, J. A. and Mead, R.",
    title = "{A Simplex Method for Function Minimization}",
    doi = "10.1093/comjnl/7.4.308",
    journal = "Comput. J.",
    volume = "7",
    pages = "308--313",
    year = "1965"
}

@article{KAGRA:2021duu,
    author = "Abbott, R. and others",
    collaboration = "KAGRA, VIRGO, LIGO Scientific",
    title = "{Population of Merging Compact Binaries Inferred Using Gravitational Waves through GWTC-3}",
    eprint = "2111.03634",
    archivePrefix = "arXiv",
    primaryClass = "astro-ph.HE",
    reportNumber = "LIGO-P2100239 ; Data release: https://zenodo.org/record/5655785, LIGO-P2100239",
    doi = "10.1103/PhysRevX.13.011048",
    journal = "Phys. Rev. X",
    volume = "13",
    number = "1",
    pages = "011048",
    year = "2023"
}

@article{Abbott:2020search,
  title={Prospects for observing and localizing gravitational-wave transients with Advanced LIGO, Advanced Virgo and KAGRA},
  author={Abbott, Benjamin P and Abbott, R and Abbott, TD and Abraham, S and Acernese, F and Ackley, K and Adams, C and Adya, VB and Affeldt, C and Agathos, M and others},
  journal={Living reviews in relativity},
  volume={23},
  pages={1--69},
  year={2020},
  publisher={Springer}
}

@article{Ashton:2018jfp,
    author = "Ashton, Gregory and others",
    title = "{BILBY: A user-friendly Bayesian inference library for gravitational-wave astronomy}",
    eprint = "1811.02042",
    archivePrefix = "arXiv",
    primaryClass = "astro-ph.IM",
    doi = "10.3847/1538-4365/ab06fc",
    journal = "Astrophys. J. Suppl.",
    volume = "241",
    number = "2",
    pages = "27",
    year = "2019"
}

@article{Smith:2019ucc,
    author = "Smith, Rory J. E. and Ashton, Gregory and Vajpeyi, Avi and Talbot, Colm",
    title = "{Massively parallel Bayesian inference for transient gravitational-wave astronomy}",
    eprint = "1909.11873",
    archivePrefix = "arXiv",
    primaryClass = "gr-qc",
    reportNumber = "LIGO Document P1900255-v1",
    doi = "10.1093/mnras/staa2483",
    journal = "Mon. Not. Roy. Astron. Soc.",
    volume = "498",
    number = "3",
    pages = "4492--4502",
    year = "2020"
}

@article{Cornish:2011ys,
    author = "Cornish, Neil and Sampson, Laura and Yunes, Nicolas and Pretorius, Frans",
    title = "{Gravitational Wave Tests of General Relativity with the Parameterized Post-Einsteinian Framework}",
    eprint = "1105.2088",
    archivePrefix = "arXiv",
    primaryClass = "gr-qc",
    doi = "10.1103/PhysRevD.84.062003",
    journal = "Phys. Rev. D",
    volume = "84",
    pages = "062003",
    year = "2011"
}

@article{Bhat:2025lri,
    author = "Bhat, Sajad A. and Tiwari, Avinash and Shaikh, Md Arif and Kapadia, Shasvath J.",
    title = "{EECT: an Eccentricity Evolution Consistency Test to distinguish eccentric gravitational-wave signals from eccentricity mimickers}",
    journal = "arXiv:2508.14850",
    eprint = "2508.14850",
    archivePrefix = "arXiv",
    primaryClass = "gr-qc",
    month = "8",
    year = "2025"
}
	
\end{document}